\documentclass[aps,prc,twocolumn,nofootinbib,amsmath,showpacs,superscriptaddress,floatfix]{revtex4}
\usepackage{graphicx,epsfig,longtable}
\usepackage{mathptmx}
\usepackage[light,first]{draftcopy} 
\usepackage{epstopdf}
\usepackage[pdftex]{hyperref}
\usepackage{wasysym}
\usepackage{multirow}
\usepackage{eqnarray,amsmath,amsbsy}
\usepackage{booktabs}
\usepackage{xcolor}

\newcommand{\beqy}{\begin{eqnarray}}
\newcommand{\eeqy}{\end{eqnarray}}
\newcommand{\bmlet}{\begin{subequations}}
\newcommand{\emlet}{\end{subequations}}
\newcounter{saveeqn}

\def\gsimeq{\,\,\raise0.14em\hbox{$>$}\kern-0.76em\lower0.28em\hbox
{$\sim$}\,\,}
\def\lsimeq{\,\,\raise0.14em\hbox{$<$}\kern-0.76em\lower0.28em\hbox
{$\sim$}\,\,}

\begin{document}

\preprint{APS/123-QED}

\title{Evolution of the $\gamma$-ray strength function in neodymium isotopes}

\author{M.~Guttormsen}
\email{magne.guttormsen@fys.uio.no}
\affiliation{Department of Physics, University of Oslo, N-0316 Oslo, Norway}
\author{K.~O.~Ay}
\affiliation{Department of Physics, Eskisehir Osmangazi University, Faculty of Science and Letters, TR-26040 Eskisehir, Turkey}
\author{M.~Ozgur}
\affiliation{Department of Physics, Eskisehir Osmangazi University, Faculty of Science and Letters, TR-26040 Eskisehir, Turkey}
\author{E.~Algin}
\affiliation{Department of Physics, Eskisehir Osmangazi University, Faculty of Science and Letters, TR-26040 Eskisehir, Turkey}
\affiliation{Department of Metallurgical and Materials Engineering, Pamukkale University, 20160 Denizli, Turkey}
\author{A. C. Larsen}
\author{F.~L.~Bello~Garrote}
\author{H.~C.~Berg}
\thanks{Present address: National Superconducting Cyclotron Laboratory, Michigan State University, East Lansing, Michigan 48824, USA}
\author{L.~Crespo Campo}
\author{T.~Dahl-Jacobsen}
\author{F.~W.~Furmyr}
\author{D.~Gjestvang}
\author{A.~G{\"o}rgen}
\author{T.~W.~Hagen}
\author{V.~W.~Ingeberg}
\affiliation{Department of Physics, University of Oslo, N-0316 Oslo, Norway}
\author{B.~V.~Kheswa}
\affiliation{Department of Physics, University of Oslo, N-0316 Oslo, Norway}
\affiliation{Department of Physics, University of Johannesburg, P.O. Box 524, Auckland Park 2006, South Africa}
\author{I.~K.~B.~Kullmann}
\affiliation{Institut d’Astronomie et d’Astrophysique, CP-226, Université Libre de Bruxelles, 1050 Brussels, Belgium}
\author{M.~Klintefjord}
\affiliation{Department of Physics, University of Oslo, N-0316 Oslo, Norway}
\author{M.~Markova}
\author{J.~E.~Midtb{\o}}
\author{V.~Modamio}
\affiliation{Department of Physics, University of Oslo, N-0316 Oslo, Norway}
\author{W.~Paulsen}
\author{L.~G.~Pedersen}
\author{T.~Renstr{\o}m}
\author{E.~Sahin}
\author{S.~Siem}
\affiliation{Department of Physics, University of Oslo, N-0316 Oslo, Norway}
\author{G.~M.~Tveten}
\affiliation{Department of Physics, University of Oslo, N-0316 Oslo, Norway}
\author{M.~Wiedeking}
\affiliation{SSC Laboratory, iThemba LABS, P.O. Box 722, Somerset West 7129, South Africa}
\affiliation{School of Physics, University of the Witwatersrand, Johannesburg 2050, South Africa}

\date{\today}
 
\begin{abstract}
The experimental $\gamma$-ray strength functions ($\gamma$SFs) of $^{142,144-151}$Nd have been studied for $\gamma$-ray energies up to the neutron separation energy using the Oslo method. The results represent a unique set of $\gamma$SFs for an isotopic chain with increasing nuclear deformation. The data reveal how the low-energy enhancement, the scissors mode and the pygmy dipole resonance evolve with nuclear deformation and mass number. This indicates that the mechanisms behind the low-energy enhancement and the scissors mode are decoupled from each other.
\end{abstract}

\maketitle


\section{Introduction}

The $\gamma$-ray strength function ($\gamma$SF) is a fruitful concept in the nuclear quasicontinuum region, describing the {\em average reduced} $\gamma$-ray transition probabilities between groups of levels. The definition was established by Bartholomew {\em et al.}~\cite{Bartholomew1973}, connecting the $\gamma$SF directly to the average partial $\gamma$-ray width of the initial levels. For $\gamma$ decay (de-excitation), the $\gamma$SF is given by
\begin{equation}
f_{\rm XL}(E_{\gamma},E_i,J_i^{\pi_i})=\frac{\langle \Gamma(E_i,J_i^{\pi_i},E_{\gamma} ) \rangle}{E_{\gamma}^{2L+1}} \rho(E_i,J_i^{\pi_i}),
\end{equation}
where the initial and final spin/parity obey the selection rules for transitions of type $X$ and multipolarity $L$. The average $\gamma$ energy is given by $E_{\gamma}=E_i-E_f$, and $\rho$ is the level density at the initial excitation energy $E_i$ with spin/parity $J_i^{\pi_i}$.

There are many experimental methods to determine the $\gamma$SF. Recently, Goriely {\em et al.}~\cite{Goriely2019} have summarized the various techniques, and have compiled experimental $\gamma$SF results in a database hosted at the IAEA~\cite{IAEA2020}.

\begin{figure*}[t]
\includegraphics[clip,width=2.1\columnwidth]{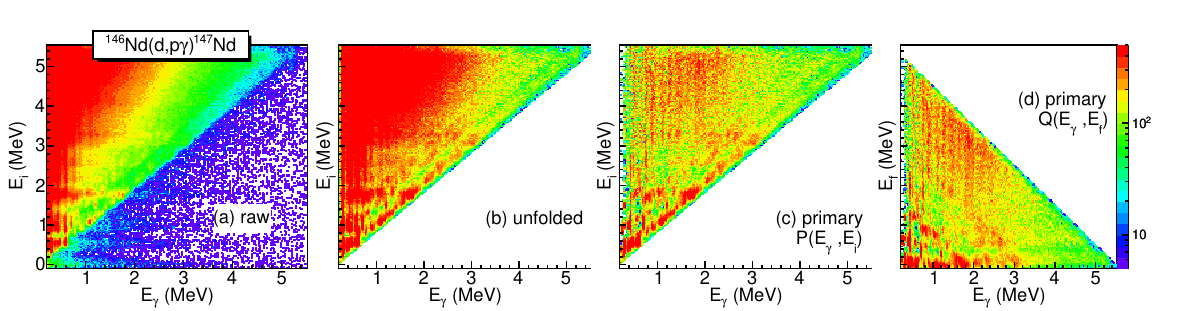}
\caption{(Color online) Proton-$\gamma$ coincidence matrices obtained in the $^{146}$Nd($d,p\gamma $)$^{147}$Nd reaction with deuteron energies of 13.5 MeV. The (a) raw, (b) unfolded, and (c) primary matrices are shown for initial excitation energies $E_i$ between the ground level and the neutron binding energy $S_n$. The primary $P(E_{\gamma},E_i)$ matrix of panel (c) represents the starting point for the Oslo method. We also show the $Q(E_{\gamma},E_f)$ matrix of panel (d), which reveals the intensities of $\gamma$-ray transitions populating final excitation energies $E_f$. All matrices have $(x,y)$-pixel sizes of $(28,31)$ keV.}
\label{fig:matrices}
\end{figure*}

The $\gamma$SF is composed of several collective modes where the dominant component is the giant electric dipole resonance (GDR) centered typically around 15~MeV of $\gamma$-ray energy. The $\gamma$SF  below the neutron separation energy ($S_n\approx 5-10$~MeV) represents the low-energy part of the GDR. As the GDR strength is $10-100$ times lower for these $\gamma$-ray energies as compared to its maximum, other smaller structures appear, such as the low-energy enhancement (LEE), the scissors mode (SM) and the pygmy dipole resonance (PDR). The magnetic spin-flip resonance is also expected at these energies, but with a negligible strength~\cite{IAEA2020}. 

Interpreting several of the $\gamma$SF structures has been a long-standing problem. Of particular interest is the nature of the LEE and SM structures and their dependence on deformation. It has been speculated~\cite{schwengner2017,naqvi2019,frauendorf2022} that these two (presumably) M1 structures are in some way connected and add up to an integrated strength, which is relatively independent on deformation and mass number. One of the experimental challenges is to measure the LEE $\gamma$SF down to the lowest $\gamma$-ray energies. Furthermore, the impact of the PDR and its interpretation as being due to neutron skin oscillations outside the $N=Z$ core are of utmost interest~\cite{savran2013}. 

 The chain of stable neodymium isotopes covers nuclei from almost-spherical to well-deformed shapes. In the present work, we have measured the $\gamma$SFs of $^{142,144-151}$Nd using particle-$\gamma$ coincidences from light-ion reactions on stable neodymium targets. The data were analyzed within the framework of the Oslo method~\cite{Gut87,Gut96,Schiller00} and the recently developed shape method~\cite{Wiedeking2021}. By also exploiting experimental results from other measurements, the evolution of the $\gamma$SF in the energy region of $E_{\gamma}\approx 0-17$~MeV and the interplay between its various components are discussed as function of deformation and mass number.

 Since both experimental and theoretical studies of the NLD in the neodymium isotopes were recently published~\cite{Gut2021}, we focus mainly on the $\gamma$SF in the present work.

The paper is outlined as follows. Section II describes the experiments at the Oslo Cyclotron Laboratory and the Oslo method. In Sect. III the spin distributions of the two applied reactions are investigated with the help of $\gamma$-ray sidefeeding intensities into the rotational ground band of $^{150}$Nd and by means of the shape method. The experimental $\gamma$SF results are presented in Sect. IV together with model fitting to the experimental data. Summary and conclusions are given in Sect.~V.

\section{Experiments and the Oslo method}

The chain of $^{142,144-151}$Nd isotopes was studied with light-ion reactions at the Oslo Cyclotron Laboratory. The targets were self-supporting metallic foils of $^{142,144,146,148,150}$Nd with thicknesses of $\approx 2$~mg/cm$^2$ and enrichments of $\approx 97$\%. The targets were bombarded with proton and deuteron beams of energies 16.0 and 13.5 MeV, respectively.

The SiRi particle-telescope system~\cite{siri} was applied to determine the outgoing particle type and energy. The 64 particle telescopes were located $\approx$~5 cm from the target in eight angles between $126^{\circ}$ and $140^{\circ}$ with respect to the beam direction. The front ($\Delta E$) and back ($E$) detectors had thicknesses of 130 and 1550 $\mu$m, respectively. The total particle energy ($E + \Delta E$) resolution was $\approx$ 150 keV (FWHM).

During the neodymium experimental campaign, the $\gamma$-detector array CACTUS~\cite{CACTUS} was replaced by the OSCAR array~\cite{Zeiser2021,Goergen2021}. The CACTUS array is equipped with 26 NaI(Tl) 5" $\times$ 5" collimated scintillator detectors at a distance of 22 cm from the target. This $\gamma$-detector array was used for the $^{144,148,150}$Nd$(d,p)$ reactions\footnote{We also analyzed the $(d, d')$ reaction, but this channel had too low of a cross section at high excitation energies.}. The new OSCAR array consists of LaBr$_3$(Ce) scintillators characterized by high efficiency combined with excellent timing and energy resolution. In the $^{142,146}$Nd$(p,p')$ and $^{146}$Nd$(d,p)$ reactions a total of 15 LaBr$_3$(Ce) detectors were placed 22 cm from the target, and in the $^{144,148,150}$Nd$(p,p')$ reactions, 30 detectors at 16 cm were used.

The Oslo method allows a simultaneous extraction of NLD and $\gamma$SF from the same particle-$\gamma$ coincidence data set. The first task is to sort the coincidence events into $\gamma$ spectra for each initial excitation energy $E_i$ of the residual nucleus. The $E_i$ value is calculated event-by-event from the detected energy of the charged ejectile and the reaction kinematics. Figure \ref{fig:matrices}(a) shows this raw $(E_{\gamma}, E_i)$ matrix of $^{147}$Nd. Each $\gamma$ spectrum is then unfolded using the known detector response functions for CACTUS or OSCAR~\cite{Gut96,om2022}. The matrix obtained from the unfolding procedure~\cite{Gut96} is displayed in Fig.~\ref{fig:matrices}(b), which presents all $\gamma$ cascades from each $E_i$. The $\gamma$ multiplicity of these statistical transitions including all $\gamma$ rays for all cascades is typically $\approx 1-4$, which can be calculated from~\cite{Gut87}
\begin{equation}
M_{\gamma}(E_i) = \frac{E_i}{\langle E_{\gamma}(E_i) \rangle},
\label{eq:energy}\\
\end{equation}
where $\langle E_{\gamma}\rangle$ is the average $\gamma$-ray energy of the $\gamma$ spectrum for an initial excitation energy $E_i$.

\begin{figure*}[t]
\begin{center}
\includegraphics[clip,width=1.8\columnwidth]{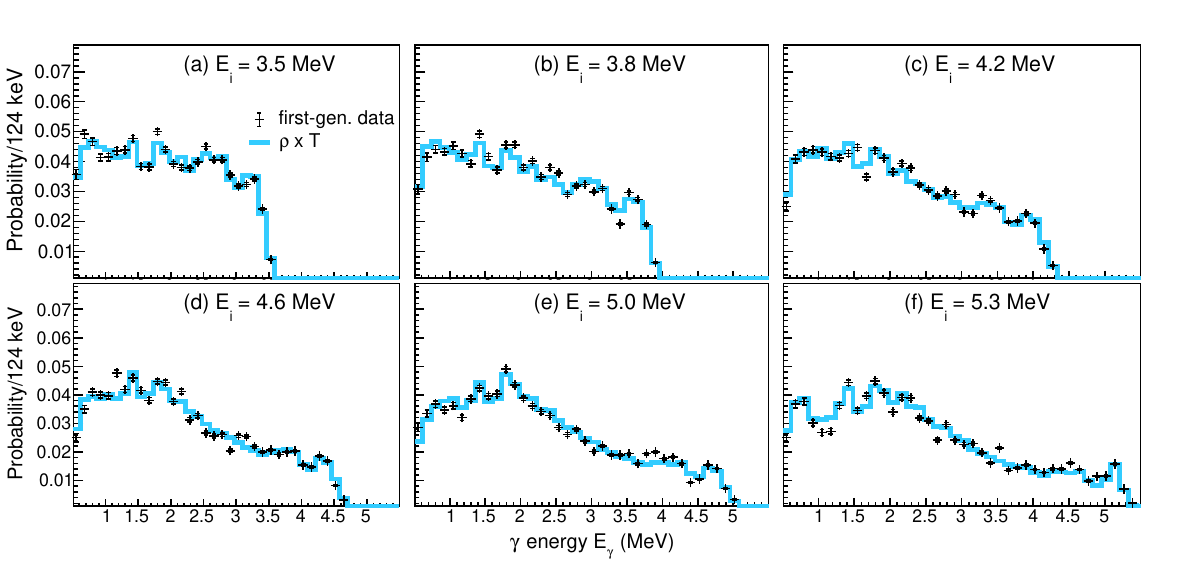}
\caption{(Color online) Probability densities of primary $\gamma$ spectra (crosses) from various initial excitation energies $E_i$ in $^{147}$Nd. The spectra are compared to the product $\rho(E_i-E_{\gamma}) {\mathcal{T}}(E_{\gamma})$ (blue histogram). The statistical error bars are less than the data crosses. Both the $\gamma$ and excitation energy dispersions are 124 keV/channel.}
\label{fig:doesitwork}
\end{center}
\end{figure*}

An important step of the Oslo method is to obtain a reliable first-generation (primary) $\gamma$-ray matrix. The construction of this matrix is based on an iterative subtraction technique~\cite{Gut87} which separates the energy distribution of the first emitted $\gamma$ rays from the distribution of higher-generation $\gamma$ rays.

Let $u_{E_i}(E_{\gamma})$ be the unfolded $\gamma$ spectrum measured at the initial excitation energy $E_i$, as shown in Fig.~\ref{fig:matrices}(b). Then, the primary spectrum at initial excitation energy $E_i$ can be obtained by subtracting a sum of weighted $u_{E_i'}(E_{\gamma})$ spectra from lower excitation energies
\begin{equation}
p_{E_i}(E_{\gamma})=u_{E_i}(E_{\gamma}) - \sum_{E_i' < E_i}w_{E_i}(E_i')u_{E_i'}(E_{\gamma}),
\end{equation}
where the weighting coefficients $w$ and first-generation spectrum $p$ are determined by an iterative procedure as described in Ref.~\cite{Gut87}. After a few iterations, the multiplicity of the primary spectrum is found to be $M_{\gamma}\approx 1$, which is compared to the higher multiplicity of the $u$ spectrum of the average total cascades. The reason for the fast convergence is the close relationship between $p$ and $w$. At a given initial excitation energy $E_i$, the functional form of $p_{E_i}(E_{\gamma})$ should end up by having the same functional form as $w_{E_i}(E_i')$ when taken as function of $E_i - E_i'$, which represents the argument $E_{\gamma}$ of $p$.

\begin{figure}[t]
    \includegraphics[clip,width=0.9\columnwidth]{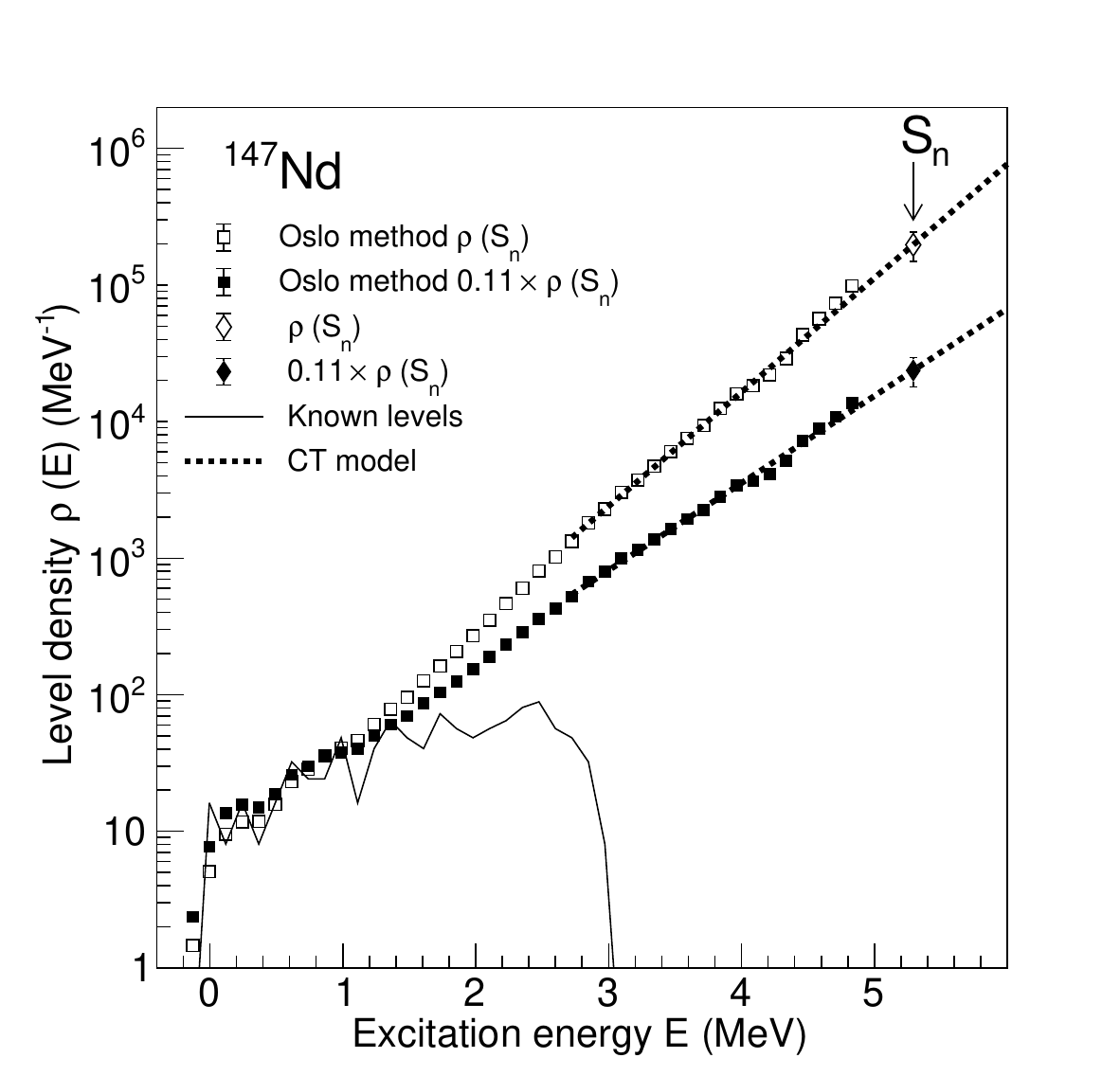}
    \caption{Level densities of $^{147}$Nd. The open and filled square data points show the results of the Oslo method applied with a total ($\rho_{\rm tot}$) and reduced ($\rho_{\rm exp}$) level density, respectively, having $\rho_{\rm exp}(S_n) = 0.11 \rho_{\rm tot}(S_n)$. The data points of the two level densities are connected to $\rho(S_n)$ (shown as diamonds) with a constant temperature (CT) model. The solid line shows the level density of known levels~\cite{NNDC}.}
    \label{fig:counting_eta}
\end{figure}

The extracted primary $P(E_{\gamma},E_i$) matrix for the $^{146}$Nd($d,p\gamma $)$^{147}$Nd reaction is shown in Fig.~\ref{fig:matrices}(c). Compared to the matrix of cascades in panel (b), we see that the intensities of $\gamma$ rays with energies at the initial excitation energy $E_i$ and slightly below, are about the same since they represent primary transitions in both matrices. On the other side, we see that the intensities of lower-energy $\gamma$ rays at high $E_i$ are strongly suppressed in the primary matrix as they represent higher-generation $\gamma$ rays. For illustration, in Fig.~\ref{fig:matrices}(d), we have also constructed the intensities of $\gamma$ rays feeding final excitation energies $E_f$ given by $Q(E_{\gamma},E_f) = P(E_{\gamma},E_i-E_{\gamma})$. The $Q$ matrix shows that $\gamma$ rays below 3 MeV decay mainly directly to four to five low-lying levels at $E_f < 200$~keV. The decay to these levels are also recognized as the lowest diagonal in the $P$ matrix.

In order to extract NLD and $\gamma$SF, we first normalize the primary spectra to unity by $\sum_{E_{\gamma}}P(E_{\gamma}, E_i)=1$. We then factorize $P$ by~\cite{Schiller00}
\begin{equation}
P(E_{\gamma}, E_i) \propto   \rho(E_i-E_{\gamma}){\cal{T}}(E_{\gamma}) ,\
\label{eqn:rhoT}
\end{equation}
where we assume that the decay probability is proportional to the NLD at the final energy $\rho(E_i-E_{\gamma})$ according to Fermi's golden rule~\cite{dirac,fermi}. The decay is also proportional to the $\gamma$-ray transmission coefficient ${\cal{T}}$, which is assumed to be independent of excitation energy according to the Brink hypothesis~\cite{brink,guttormsen2016}.

Provided that relation (\ref{eqn:rhoT}) holds, we extract the one-dimensional vectors $\rho$ and ${\cal{T}}$ from the two-dimensional $P$ matrix using the least $\chi ^2$ fit iteration procedure of Schiller {\em et al.}~\cite{Schiller00}. For a successful fit we adopt a part of the $P$ matrix where the decay can be considered statistical. A guide to this qualitative statement is to monitor at what energy region the first-generation method works well, meaning that we obtain the expected multiplicity of $M_{\gamma}\approx 1$. In the case of $^{147}$Nd in Fig.~\ref{fig:doesitwork}, the energy region of 3.0~$< E_i < $~5.4~MeV with $E_{\gamma}> 0.5$~MeV works well as confirmed by the agreement between experimental data points and the results (blue histogram) using the fit functions $\rho$ and ${\cal T}$. The figure includes six out of the 21 available $\gamma$ spectra that have been used to determine $\rho$ and ${\cal{T}}$ in the fitting procedure. 

With the transmission coefficient ${\cal{T}}$ in hand, we obtain the $\gamma$SF by~\cite{kopecky1990}
\begin{equation}
f (E_{\gamma}) =\frac{1}{2 \pi}\frac{{\cal {T}}(E_{\gamma})}{ E_{\gamma}^3 },
\end{equation}
as dipole transitions dominate the decay in the quasicontinuum \cite{Capote09}.
 
The local variation of data points of $\rho$ and $\cal{T}$ are uniquely determined through the fit, but the scale and slope of these functions are still undetermined. It has been shown that transformations of the type~\cite{Schiller00}
\begin{eqnarray}
\tilde{\rho}(E-E_\gamma)&=&A\exp[\alpha(E-E_\gamma)]\,\rho(E-E_\gamma),
\label{eq:array1}\\
\tilde{{\mathcal{T}}}(E_\gamma)&=&B\exp(\alpha E_\gamma){\mathcal{T}} (E_\gamma),
\label{eq:array2}
\end{eqnarray}
give identical fits to the primary $\gamma$ spectra. Therefore, $A$, $\alpha$ and $B$ are parameters that have to be determined from other experimental data or systematics.

\begin{table*}[bth]
\centering
\caption{The quadrupole deformation $\beta_2$ and parameters for extracting experimental NLD and $\gamma$SF.}
\begin{tabular}{cccccccccc}
\hline
\hline
Nucleus    &$\beta_2$   &$T_{\rm CT}$&$E_d$&$\sigma_d$  &$S_n$& $\sigma(S_n)$ &$D_0$          & $\rho(S_n)$      &$\langle\Gamma_{\gamma}\rangle$\\
           &            & (MeV) &     (MeV)&               &(MeV)& RMI           & (eV)          &(10$^6$MeV$^{-1})$&   (meV)            \\
\hline \\
$^{142}$Nd &0.092(2)    &0.65(5)&     2.5  & 3.0  &9.828 &  6.6           &19(4)$^a$      &    1.23(35)$^b$   &    77(20)$^c$      \\
$^{144}$Nd &0.125(2)    &0.63(3)&     2.5  & 2.8  &7.817 &  6.3           &37.6(21)       &    0.32(5)        &    74.2(18)        \\
$^{145}$Nd &0.138(5)$^d$&0.59(3)&     1.3  & 2.9  &5.755 &  5.9           &450(50)        &    0.16(4)        &    51(4)           \\
$^{146}$Nd &0.151(2)    &0.62(3)&     1.5  & 2.6  &7.565 &  6.2           &17.8(7)        &    0.67(11)       &    74(3)           \\
$^{147}$Nd &0.176(5)$^d$&0.57(3)&     0.5  & 2.0  &5.292 &  5.8           &346(50)        &    0.20(5)        &    54(4)           \\
$^{148}$Nd &0.200(2)    &0.59(3)&     1.4  & 2.5  &7.333 &  6.1           &5.9(11)        &    2.4(6)         &    68.8(60)        \\
$^{149}$Nd &0.242(5)$^d$&0.54(3)&     0.5  & 2.3  &5.039 &  5.8           &165(14)        &    0.42(9)        &    45(3)           \\
$^{150}$Nd &0.283(2)    &0.61(4)&     1.2  & 2.9  &7.376 &  6.2           &3.0(10)$^a$    &    4.8(18)$^b$    &    70(20)$^c$      \\
$^{151}$Nd &0.314(10)$^d$&0.54(3)&    0.4  & 2.6  &5.335 &  6.0            &169(11)       &    0.43(9)        &    67(25)          \\
\hline
\hline
\end{tabular}
\label{tab:gsf_parameters}

$^a$Adjusted to reproduce $\rho(S_n)$. $^b$Estimated from systematics~\cite{Gut2021}. \\ $^c$Estimated from $^{144,146,148}$Nd. $^d$Interpolated between even-mass neighbors. 

\end{table*}

At low excitation energies, we normalize the NLD to known discrete levels~\cite{NNDC}. At high excitation energies, we use the measured average neutron $s$-wave resonance spacing $D_0$~\cite{MugAtlas} at the neutron separation energy $S_n$. To convert the measured $D_0$ to the total level density, we insert $E=S_n$ into the spin distribution~\cite{Ericson59}
\begin{equation}
g(E,J) \simeq \frac{2J+1}{2\sigma^2(E)}\exp\left[-(J+1/2)^2/2\sigma^2(E)\right],
\label{eq:spindist}
\end{equation}
where $J$ is the spin quantum number. The function of the spin cutoff parameter is given by~\cite{Capote09}
\begin{equation}
\sigma^2(E)=\sigma_d^2 + \frac{\sigma^2(S_n)-\sigma_d^2}{S_n-E_d}\left(E-E_d\right),
\label{eq:sigE}
\end{equation}
where $\sigma_d^2$ is determined from known discrete levels at low excitation energy $E=E_d$ and $\sigma^2(S_n)$ is determined from the rigid-body moment of inertia (RMI) estimate, as shown in our previous work~\cite{Gut2021}.

Table~\ref{tab:gsf_parameters} lists the quadrupole deformations and parameters needed for extracting $\rho(S_n)$. The $\beta_2$ values of the even-mass isotopes are taken from the compilation of Pritychenko {\em et al.}~\cite{pritychenko2016}. For the odd-mass isotopes, we assume a deformation that is the average of their even-mass neighbors. The table also includes the temperature $T_{\rm CT}$ extracted by a $\chi ^2$ fit of the constant-temperature (CT) formula~\cite{Ericson59}
\begin{equation}
\rho_{\text{CT}}(E)=(1/T_{\rm CT}) \exp{[(E-E_0)/T_{\text{CT}}]}
\label{eq:ct}
\end{equation}
to the experimental high-energy data points and the predicted $\rho(S_n)$ value of Table~\ref{tab:gsf_parameters}. 
Such a fit is shown in Fig.~\ref{fig:counting_eta} where $\rho_{\text{CT}}$ was fitted to the data points in the excitation region of $E=2.7 - 4.1$~MeV. The energy shift parameter is given by $E_0=S_n -T_{\text{CT}}\ln[T_{\text{CT}} \; \rho(S_n)]$. 
Further details on the extraction of NLDs in the neodymiums are given in our previous work~\cite{Gut2021}. 
For convenience, we list the NLD parameters from Ref.~\cite{Gut2021}in Table~\ref{tab:gsf_parameters}.

The last column of Table~\ref{tab:gsf_parameters} lists the average $\gamma$ widths for $\ell = 0$ neutron capture reactions compiled in Ref.~\cite{MugAtlas}. These quantities are exploited to determine the scaling of ${\cal {T}}$ (parameter $B$ of Eq.~(\ref{eq:array2})) by reproducing the average, experimental $\gamma$-decay width~\cite{Schiller00,voin1}
\begin{eqnarray}
\langle\Gamma_{\gamma} (S_n)\rangle=\frac{1}{2\pi\rho(S_n, J_i, \pi)} \sum_{J_f}&&\int_0^{S_n}{\mathrm{d}}E_{\gamma} {\cal {T}}(E_{\gamma})
\nonumber\\
&&\times \rho(S_n-E_{\gamma}, J_f),
\label{eq:norm}
\end{eqnarray}
where the summation and integration run over all final levels with spin $J_f$ that are accessible from initial spin $J_i$ by $E1$ or $M1$ transitions with energy $E_{\gamma}$. 
The integral is performed with the measured experimental $\rho$ and $\cal T$ data points, however, in the case of missing data points at the lowest and highest energies, extrapolations are used.
The Oslo method has been extensively tested and discussed by Larsen {\em et al.}~\cite{Lars11}. The Oslo method software is available on the Oslo Cyclotron Laboratory GitHub~\cite{om2022}.

\section{Spin distributions of the applied reactions}

Before we can extract the $\gamma$SFs from the first-generation matrix, we have to consider the spin distribution populated in these light-ion reactions compared to the intrinsic spin distribution of the heavy nuclei studied. Table~\ref{tab:gsf_parameters} shows typically $\sigma(S_n) \approx 6$, which represents (see Eq.~(\ref{eq:spindist})) an average spin of $\langle J \rangle \approx 7$ with a  negligible contribution of spins above $J \approx 15-20$. 
This is a significantly larger spin distribution than expected from the applied $(p,p')$ and $(d,p)$ reactions at low beam energies\footnote{For previous helium-induced reactions on lighter nuclei, the populated spin distribution was much closer to the real intrinsic spin distribution, and performing corrections was not necessary.}.

In the following, we aim to estimate experimentally the ratio
\begin{equation}
  \eta= \frac{\rho_{\rm exp}(S_n)}{\rho_{\rm tot}(S_n)}
\end{equation}
between the level density populated in the reaction and the total, intrinsic level density\footnote{By intrinsic level density we mean all available levels within an excitation energy bin (independent on the nuclear reaction applied).} given in Table~\ref{tab:gsf_parameters}. With this reduction factor, we can estimate the experimental level density $\rho_{\rm exp}(E)$ for excitation energies up to the neutron separation energy $S_n$, as demonstrated in Fig.~\ref{fig:counting_eta} for $^{147}$Nd by the filled black data points.

It is obvious that the observed level density has a less steep slope compared to that without a reduction of available spins. Since the observed $P(E_{\gamma}, E_i)$ matrix represents the experimental spin range, $\rho_{\rm exp}(E)$ must be adopted in Eq.~(\ref{eqn:rhoT}). By replacing $\rho_{\rm tot}(E)$ with $\rho_{\rm exp}(E)$, the slope of ${\cal{T}}(E_{\gamma})$ will correspondingly change to fit the observed $P(E_{\gamma}, E_i)$ landscape. As seen from Eqs.~(\ref{eq:array1}) and (\ref{eq:array2}), a less steep slope of $\rho$ will induce a less steep slope of ${\cal{T}}$ as well.

In the following, we adopt two techniques to estimate $\eta$, namely ({\em i}) the sidefeeding into the rotational ground-state band for well-deformed nuclei and ({\em ii}) the recently developed shape method~\cite{Wiedeking2021}. Only $^{150}$Nd works for the sidefeeding method, whereas the shape method may be used for all nuclei if the final levels are known and experimentally separable.

\begin{figure}[ht]
\begin{center}
\includegraphics[clip,width=1.0\columnwidth]{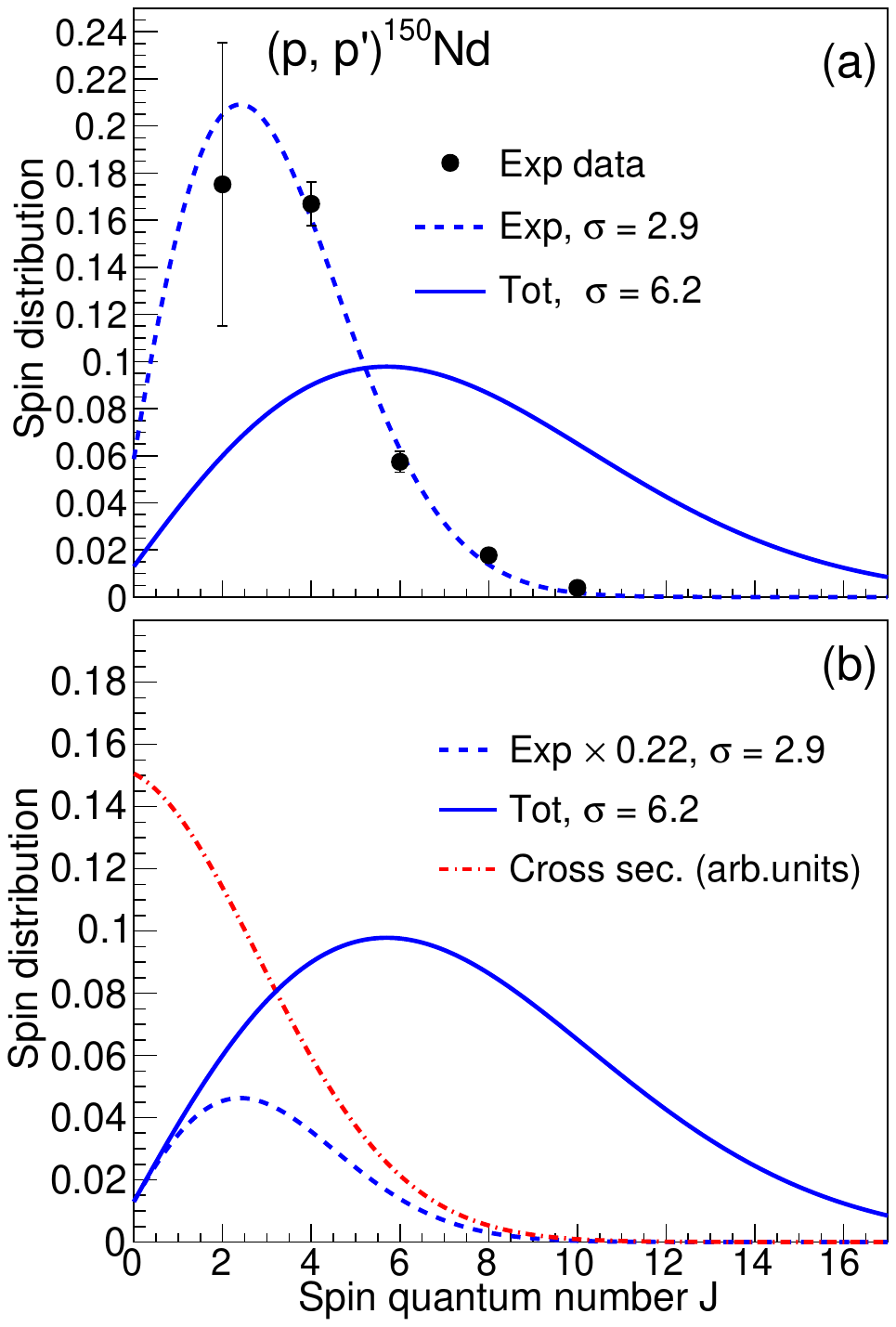}
\caption{(Color online) Spin distributions in $^{150}$Nd. Panel (a) shows the intrinsic spin distribution (solid curve) with a spin cutoff parameter of $\sigma=6.2$. The experimental data points are the $S(J)$ values obtained by the evaluated sidefeeding, which are fitted to a spin distribution (dashed curve) with $\sigma=2.9$. Panel (b) shows the experimental spin distribution (dashed curve) with $\sigma=2.9$ normalized to the lowest spin of the intrinsic spin distribution. The dashed-dotted curve (arbitrary units) shows the ratio between the experimental and total spin distributions, and will be referred to as the average probability $p^{\rm level}(E_i,J_i)$ of populating individual levels of spin $J_i$, where the value at $J = 0$ is normalized to unity.}
\label{fig:spindistribution}
\end{center}
\end{figure}

\subsection{Spin distribution of the $(p,p')$ reaction}

The $^{150}$Nd isotope is a well-behaving rotor with a quadrupole deformation of $\beta_2 = 0.283$~\cite{pritychenko2016}. With an initial excitation energy gate of $E_i=7.2-7.8$ MeV, we have evaluated the efficiency-corrected ground-state band $\gamma$-ray intensities $I_{\gamma}(J\rightarrow J-2)$ and from these values estimated the sidefeeding of spin $J$ from the quasicontinuum by
\begin{equation}
S(J)=I_{\gamma}(J\rightarrow J-2) - I_{\gamma}(J+2\rightarrow J).
\end{equation}
The intensities in the ground-state band fade exponentially with spin, and the highest transition found was the $10^+ \rightarrow 8^+$ 468.9-keV $\gamma$-ray line. Assuming that this $10^+$ level also collects the decay from higher spins, we fit the sidefeeding spin distribution to Eq.~(\ref{eq:spindist}) with the spin cutoff parameter as a free parameter. The fit result for the experimental data is shown in Fig.~\ref{fig:spindistribution}(a), with $\sigma_{\rm exp}=2.9(2)$. For comparison, also the intrinsic spin distribution with $\sigma_{\rm tot}=6.2$ is displayed as a solid curve.

The two spin distributions $g(S_n,J)_{tot}$ and $g(S_n,J)_{exp}$ of Fig.~\ref{fig:spindistribution}(a) are normalized to unity by integrating Eq.~(\ref{eq:spindist})) for all $J$.
However, with the assumption that there is no spin reduction for the lowest spin, \textit{i.e.}, $J = 0$ for $^{150}$Nd, we find the level-density reduction factor by
\begin{equation}
\eta = \frac{g(S_n,J=0)_{\rm{tot}}}{g(S_n,J=0)_{\rm{exp}}}=0.22(2).
\label{eq:gSn0}
\end{equation}
 The function $\eta g(S_n,J)_{\rm{exp}}$ is shown in Fig.~\ref{fig:spindistribution}(b), which coincides with the lowest spins of the $g(S_n,J)_{\rm{tot}}$ distribution.
 We have calculated the side feeding for three excitation-energy bins: 5.1, 6.3 and 7.5 MeV and found that the populated spin distribution varies with less than 10\%. Therefore, we have assumed that this populated spin distribution is also valid at the neutron separation energy to estimate the reduction factor $\eta$ at $S_n$.

The average probability for populating individual levels of spin $J_i$ at excitation energy $E_i$ is given by
\begin{equation}
p^{\rm level} (E_i, J_i) \propto \frac{g(E_i, J_i)_{\rm{exp}}}{g(E_i, J_i)_{\rm{tot}}},
\label{eq:xsect}
\end{equation}
which is shown as the red dashed-dotted curve (arbitrary units) in Fig.~\ref{fig:spindistribution}(b). The probability $p^{\rm level} (E_i, J_i)$ has to be taken into account in the shape method if $\gamma$-decay rates to final levels with different spins are compared. The population probabilities for various initial spins are listed in Table~\ref{tab:xsections}. It is interesting to note from Eqs.~(\ref{eq:spindist}) and (\ref{eq:xsect}) that the functional form of $p^{\rm level}$ for the present surface-induced light-ion reactions follows the right part of a Gaussian centered at $J=-1/2$, see red dashed-dotted curve in Fig.~\ref{fig:spindistribution}(b).

\begin{table}[bth]
\centering
\caption{Average probability $p^{\rm level}$ for populating individual levels in the $(p,p')$ and $(d,p)$ reactions at $S_n$. The data are normalized to unity for $J_i=0$ or $1/2$.}
\begin{tabular}{cc|cc}
\hline
\hline
    \multicolumn{2}{c|}{$^{150}$Nd$(p,p')^{150}$Nd}& \multicolumn{2}{c}{$^{144}$Nd$(d,p)^{145}$Nd}\\
    \hline
 $J_i$ & $p^{\rm level} $ & $J_i$ & $p^{\rm level} $ \\
\hline
       0    &1.00   &   1/2& 1.00   \\
       1    &0.91   &   3/2& 0.72   \\
       2    &0.76   &   5/2& 0.41   \\
       3    &0.57   &   7/2& 0.19   \\
       4    &0.39   &   9/2& 0.07   \\
       5    &0.25   &  11/2& 0.02   \\
       6    &0.14   &  13/2& 0.00   \\
       7    &0.07   &  15/2& 0.00   \\
       8    &0.04   &  17/2& 0.00   \\
       9    &0.02   &  19/2& 0.00   \\
       10   &0.01   &  21/2& 0.00   \\
\hline
\hline
\end{tabular}
\label{tab:xsections}
\end{table}

\begin{figure*}[t]
\begin{center}
\includegraphics[clip,width=1.9\columnwidth]{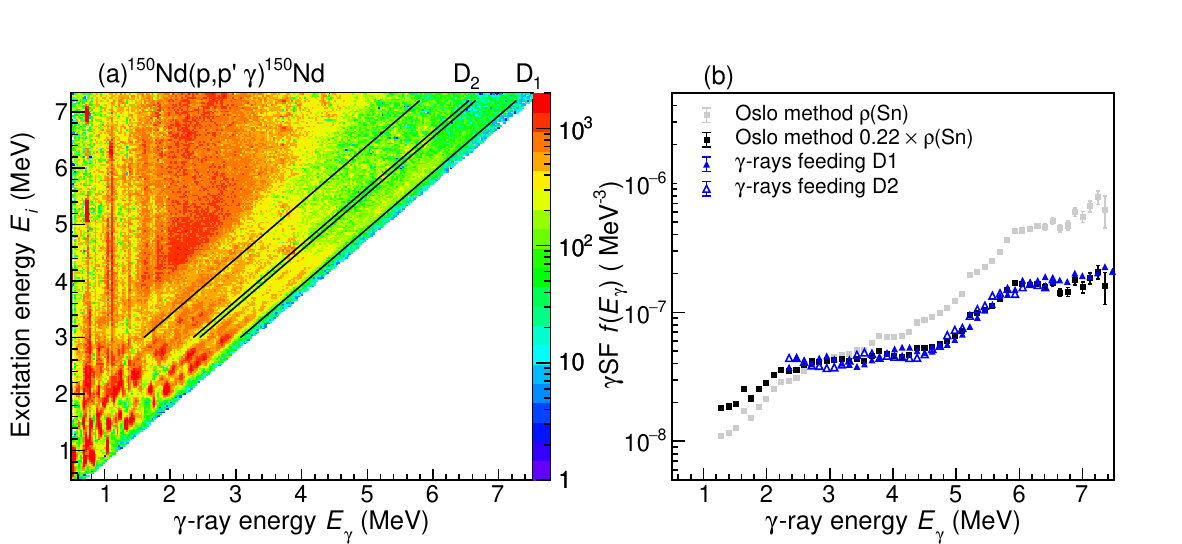}
\caption{(Color online) (a) The primary $\gamma$ matrix $P(E_{\gamma}, E_i)$ of $^{150}$Nd showing the cuts for the two diagonals. (b) The resulting $\gamma$SF from the shape method (filled and open blue triangles) compared to the Oslo method using $\eta = 1.0$ (solid grey squares) and $\eta = 0.22$ (solid black squares) from the side-feeding technique.}
    \label{fig:shape150}
\end{center}
\end{figure*}

We will now test if the above findings are consistent with the shape method results and known $(n,\gamma)$ data. The shape method~\cite{Wiedeking2021} relies on measuring the number of counts $N_D$ in a diagonal $D$ of the primary matrix defined by $E_i=E_{\gamma}+E_f$ for a fixed value of $E_f$. 
Examples of such diagonals are revealed as $D1$ and $D2$ in Fig.~\ref{fig:shape150}(a).
The number of $\gamma$-ray transitions with energy $E_{\gamma}$ from a given initial excitation energy $E_i$, is given by
\begin{equation}
 N_D\propto f(E_{\gamma})E_{\gamma}^3 \sum_{[J_f]}\sum_{J_i=J_f-1} ^{J_i=J_f+1}p^{\rm level }(E_i,J_i) \;g (E_i,J_i)_{\rm tot},
 \label{eq:ndshape}
\end{equation}
where the fixed final excitation energy $E_f= E_i- E_{\gamma}$ defines the diagonal $D$. All transitions are assumed to be dipole as the dipole strength is known to be dominant within the quasicontinuum~\cite{Capote09}. The notation $[J_f]$ describes the spins of the final levels within the diagonal, e.g., if the diagonal contains four levels with $[J_f]$, then $\sum_{[J_f]}$ is the sum over those corresponding four terms. The second sum is restricted to the available spins $J$ populated by dipole transitions connecting initial and final levels, which generally include three initial spins. However, in the case of $J_f=0$, only the $J_i=1$ spin is included and for $J_f=1/2$, only the $J_i=1/2$ and $J_i=3/2$ spins are included.

\begin{figure*}[t]
\begin{center}
\includegraphics[clip,width=1.9\columnwidth]{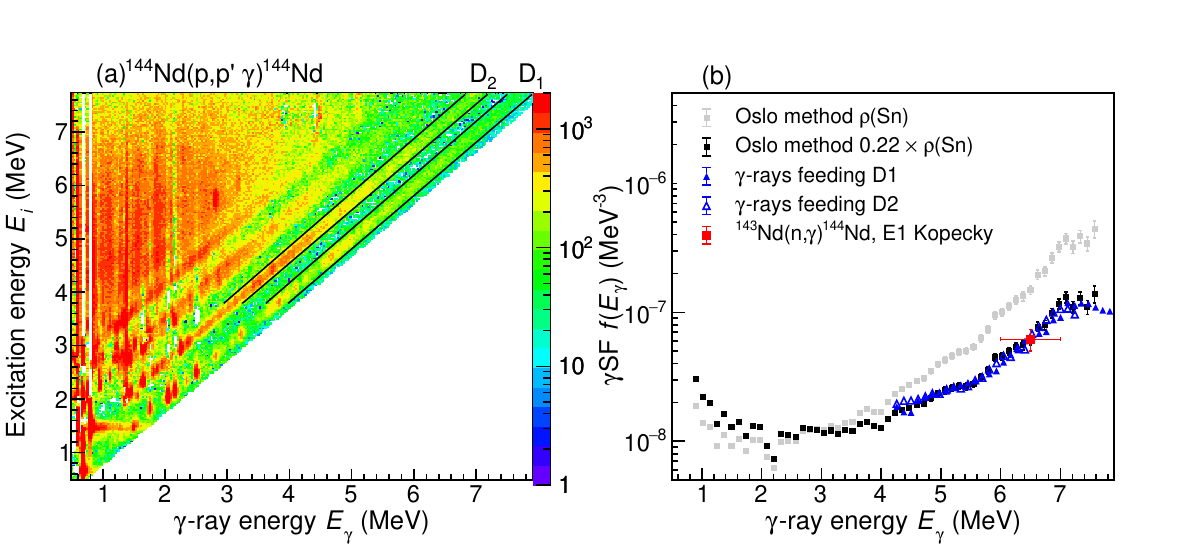}
\caption{(Color online) (a) The primary $\gamma$ matrix $P(E_{\gamma}, E_i)$  of $^{144}$Nd showing the cuts for the two the diagonals.  (b) The resulting $\gamma$SF from the shape method (filled and open blue triangles) compared to the Oslo method using $\eta = 1.0$ and $0.22$. The filled red square data point is taken from discrete resonance capture data (DRC)~\cite{IAEA2020}.}
    \label{fig:shape144}
\end{center}
\end{figure*}

\begin{figure*}[t]
\begin{center}
\includegraphics[clip,width=1.9\columnwidth]{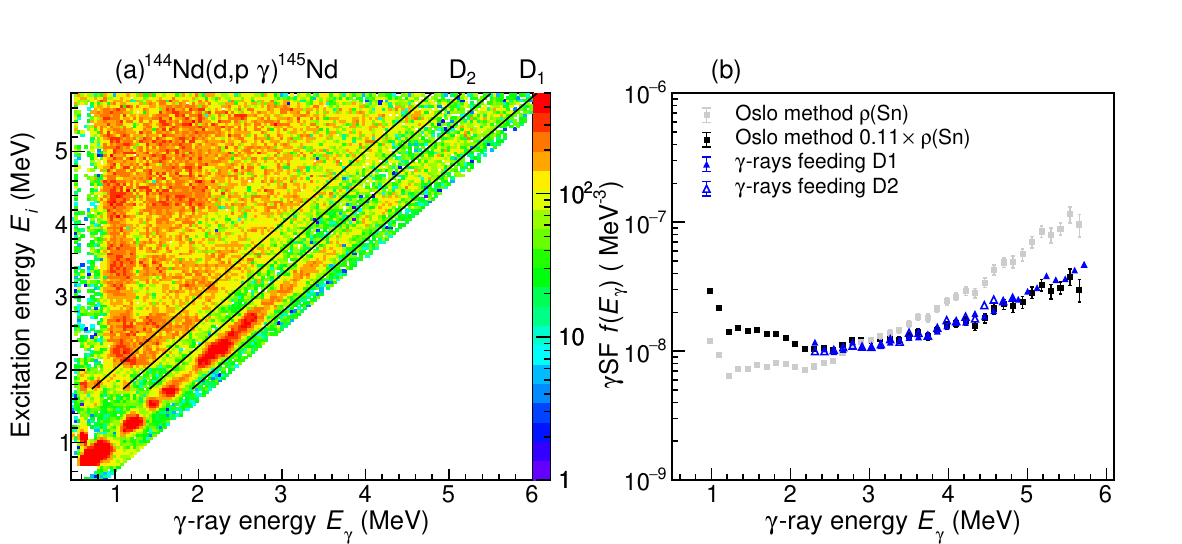}
\caption{(Color online) (a) The primary $\gamma$ matrix $P(E_{\gamma}, E_i)$ of $^{145}$Nd showing the cuts for the two diagonals. (b) The resulting $\gamma$SF from the shape method (filled and open blue triangles) compared to the Oslo method using $\eta = 1.0$ and $0.11$.}
    \label{fig:shape145}
\end{center}
\end{figure*}

The primary $P(E_{\gamma}, E_i)$ matrix for $^{150}$Nd is shown in Fig.~\ref{fig:shape150}(a) including two diagonals with their integration limits shown as black lines. Diagonal $D_1$ includes the 0$^+$ (0 keV), 2$^+$ (130 keV) and 4$^+$ (381 keV) final levels and diagonal $D_2$ includes 12 levels in the final excitation energy region $0.85 - 1.3$~MeV with average spin of $\langle J_f \rangle \approx 3.0$. 

The shape method implemented in the present work is based on the code {\sf diablo.c} available on the Oslo Cyclotron GitHub~\cite{om2022}. The algorithm of the code steps through one initial excitation energy $E_i$ of the primary $P(E_{\gamma},E_i)$ matrix and integrates the number of counts at the diagonals $D_1$ and $D_2$ within the window $E_i \pm \Delta E_i/2$, $\Delta E_i$ being the bin size. From the counts $N_{D1}$ and $N_{D2}$, a pair of internally normalized values $f(E_{\gamma 1})$ and $f(E_{\gamma 2})$ is extracted by exploiting the proportionality of Eq.~(\ref{eq:ndshape}). These pairs are then connected together by a sewing technique based on logarithmic interpolation. The obtained strength function $f(E_{\gamma})$ has in principle the correct functional form, but the absolute normalization is arbitrary and must be determined by other means. More details of the shape method are given by Wiedeking {\em et al.}~\cite{Wiedeking2021}.

Figure~\ref{fig:shape150}(b) shows the result of the shape method giving a perfect overlap with the Oslo method data using a level-density reduction factor $\eta = 0.22$ at $S_n$. Here, the shape-method data points are multiplied by a common absolute-normalization factor, which is found by a $\chi^2$ fit to the Oslo data in the $E_{\gamma}=2.5-7.3$~MeV energy region. The $\gamma$SF data points from populating the two diagonals (filled and open blue triangles) scatter slightly indicating that the systematic uncertainties with the shape method is small in the case of $^{150}$Nd. The fact that the sidefeeding technique and the shape method give consistent results is very gratifying.

We also test the shape method on $^{144}$Nd where known $(n,\gamma)$ data exist for comparison. Figure~\ref{fig:shape144}(a) shows that another advantage with this almost spherical nucleus ($\beta_2=0.125$) is that the diagonal to the 0$^+$ (0 keV) and 2$^+$ (697 keV) are well separated and thus more accurate integrals for $D_1$ and $D_2$ can be obtained. The 0$^+$ ground level is reached by dipole transitions from initial spin/parities 1$^{\pm}$, whereas the 2$^+$ level is populated by decay from the 1$^{\pm}$, 2$^{\pm}$ and 3$^{\pm}$. It is therefore important to use reasonable probabilities $p^{\rm level}$ (see Table \ref{tab:xsections}) for the initial spins populated in the reaction\footnote{If the two diagonals represent decay to levels with identical spin-parities or with a broad range of spin-parities, the values of $p^{\rm level}$ can be kept fixed for all spins.}.

The results of the shape method are displayed in Fig.~\ref{fig:shape144}(b). Again we see a good agreement between the $\gamma$SF from the shape method and the Oslo method using $\eta = 0.22$. In addition, the two $\gamma$SFs agree well with the discrete resonance capture data (DRC)~\cite{IAEA2020}, which gives additional support to our procedure described above.

We conclude that the two test cases $^{144,150}$Nd strongly suggest that a common level density reduction factor of $\eta = 0.22(2)$ at $S_n$ is reasonable for the $(p, p')$ reaction with 16-MeV protons on these neodymium isotopes.

\begin{figure}[t]
\begin{center}
\includegraphics[clip,width=1.\columnwidth]{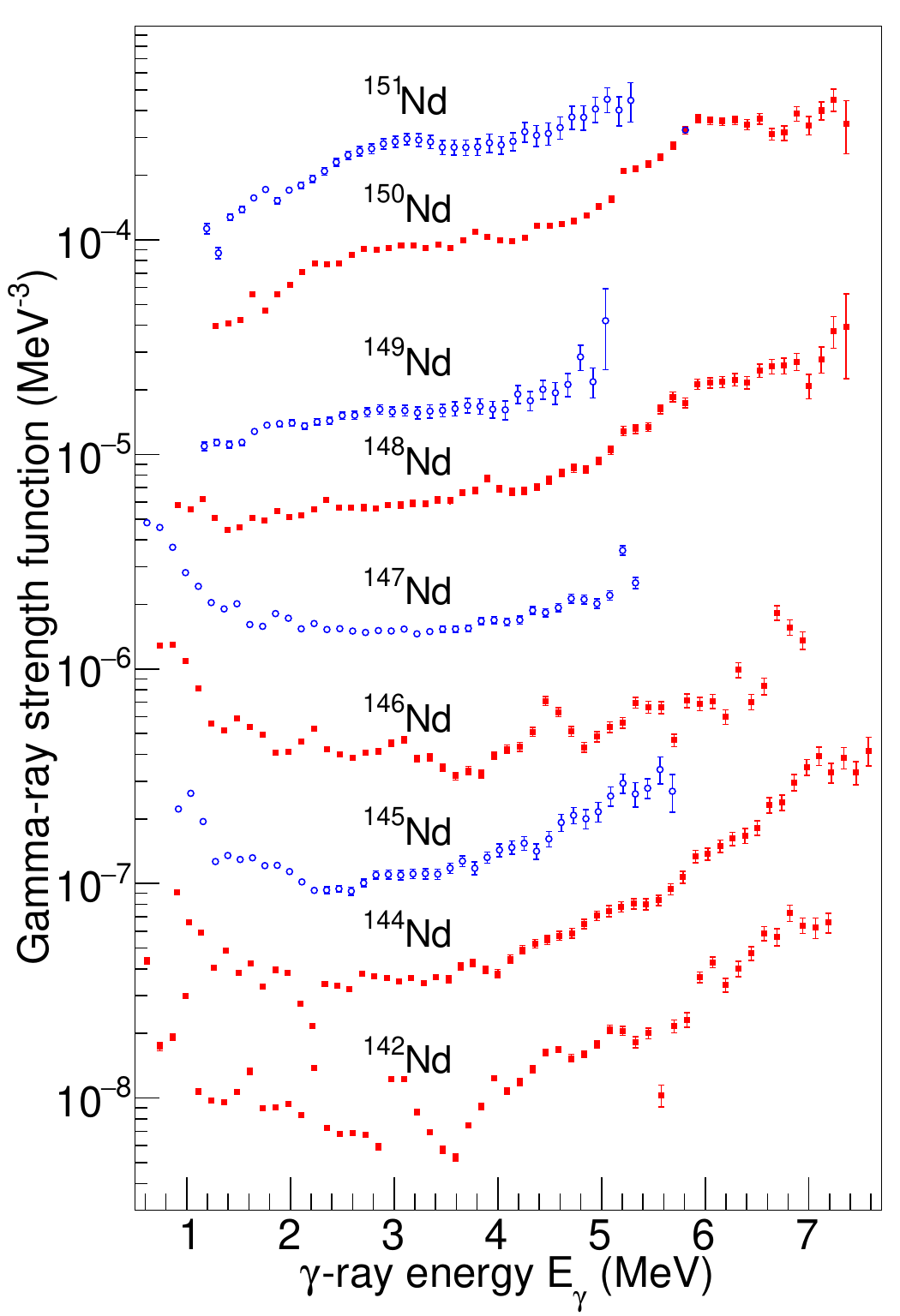}
\caption{(Color online) The $\gamma$SFs with spin reduction factors of ${\eta= 0.22}$ and $0.11$ for $^{142,144,146,148,150}$Nd (red symbols) and $^{145,147,149,151}$Nd (blue symbols), respectively (see text). Except for $^{142}$Nd, the $\gamma$SFs are separated by multiplying the next data set with a factor of 3 for better visualization.}
    \label{fig:allinone}
\end{center}
\end{figure}

\subsection{Spin distribution of the $(d,p)$ reaction}

In the present work, $(d,p)$ reactions are used to study the odd-$A$ neodymium isotopes. Here, the sidefeeding method cannot be applied due to many close-lying $\gamma$-ray lines that are not separated due to the limited detector resolution. Also, intraband transitions connecting close-lying rotational bands complicate the extraction of sidefeeding from the quasicontinuum. However, the shape method is applicable provided that the two diagonals include levels of known spin/parities.

The best case for the shape method applied to the $(d,p)$ reaction is $^{145}$Nd, where the lowest diagonal $D_1$ is well defined with the levels 7/2$^-$ (0 keV), 3/2$^-$ (67 keV) and 5/2$^-$ (73 keV). Diagonal $D_2$ is more problematic, however, we have taken 10 levels in the final excitation region $0.66 - 1.09$~MeV with average spin of $\langle J_f \rangle \approx 2.9$.

Figure \ref{fig:shape145}(a) shows the diagonals and integration limits for $^{145}$Nd, and the shape method results are displayed as filled and open blue triangles in Fig.~\ref{fig:shape145}(b). As shown, the Oslo method with intrinsic spin distribution (solid grey squares) exhibits a  $\gamma$SF too steep compared to the shape method. By introducing a level density reduction factor at $S_n$ of $\eta=0.11(2)$, a very good overlap between the Oslo and shape method is obtained. As for the $(p, p')$ reaction, we assume that the experimental spin distribution follows Eq.~(\ref{eq:spindist}). Again we can estimate the probability $p^{\rm level} (E_i, J_i)$ for populating individual levels of spin $J_i$ at excitation energy $E_i$ using Eq.~(\ref{eq:xsect}). Table~\ref{tab:xsections} lists the ($d,p$) spin population probabilities $p^{\rm level}$ at $S_n$, which are normalized to unity for $J=1/2$.

Figure~\ref{fig:allinone} summarizes the results obtained by introducing spin corrections in the Oslo method. With reduction factors of $\eta=0.22(2)$ and $\eta=0.11(2)$ for the $(p, p')$ and $(d,p)$ reactions, respectively, the $\gamma$SFs follow a systematic trend from isotope to isotope. This feature is encouraging and indicates that our corrections are sound.

We should mention that the extracted data points at or below $E_{\gamma}\approx 1$~MeV of Fig.~\ref{fig:allinone} may have been distorted by an imperfect subtraction of strong $\gamma$ lines in the first-generation procedure. Such structures appear as vertical ridges and/or valleys in some of the primary Nd matrices, and therefore data points at or below $E_{\gamma}\approx 1$~MeV should be taken with caution.

\section{Composition and evolution of the $\gamma$SF}

 The $\gamma$SF is composed of several structures which interplay and add up to the total $\gamma$SF. Many of these structures depend strongly on the quadrupole deformation $\beta_2$, which makes the chain of neodymiums isotopes of particular interest. The giant dipole resonance (GDR) is known to split into two components with deformation and the pygmy dipole resonance (PDR) is expected to be stronger with increasing neutron excess. Furthermore, the scissors mode (SM) strongly depends on deformation and finally the low-energy enhancement (LEE) seems to be absent for deformed, heavy systems. 

By introducing semi-empirical models~\cite{Capote09,Goriely2019} for all these structures, a total of 18 parameters have to be determined. This complicates a simultaneous fit to the data, and the following fitting strategy is chosen. Since the GDR is well separated from other structures around $\approx 15$~MeV, we will first fit this part of the GDR. Using these parameters, the low-energy E1 tail from the GDR is established, and we may introduce the weaker and lower-lying structures on top of this. In cases where no experimental data exist for certain $\gamma$-energy regions, we use the neighboring isotopes as a guidance. We apply the fit method implemented in ROOT, which is based on the Minuit package~\cite{james1981} with Hessian matrix error analysis.

\begin{figure*}[t]
\begin{center}
\includegraphics[clip,width=2.0\columnwidth]{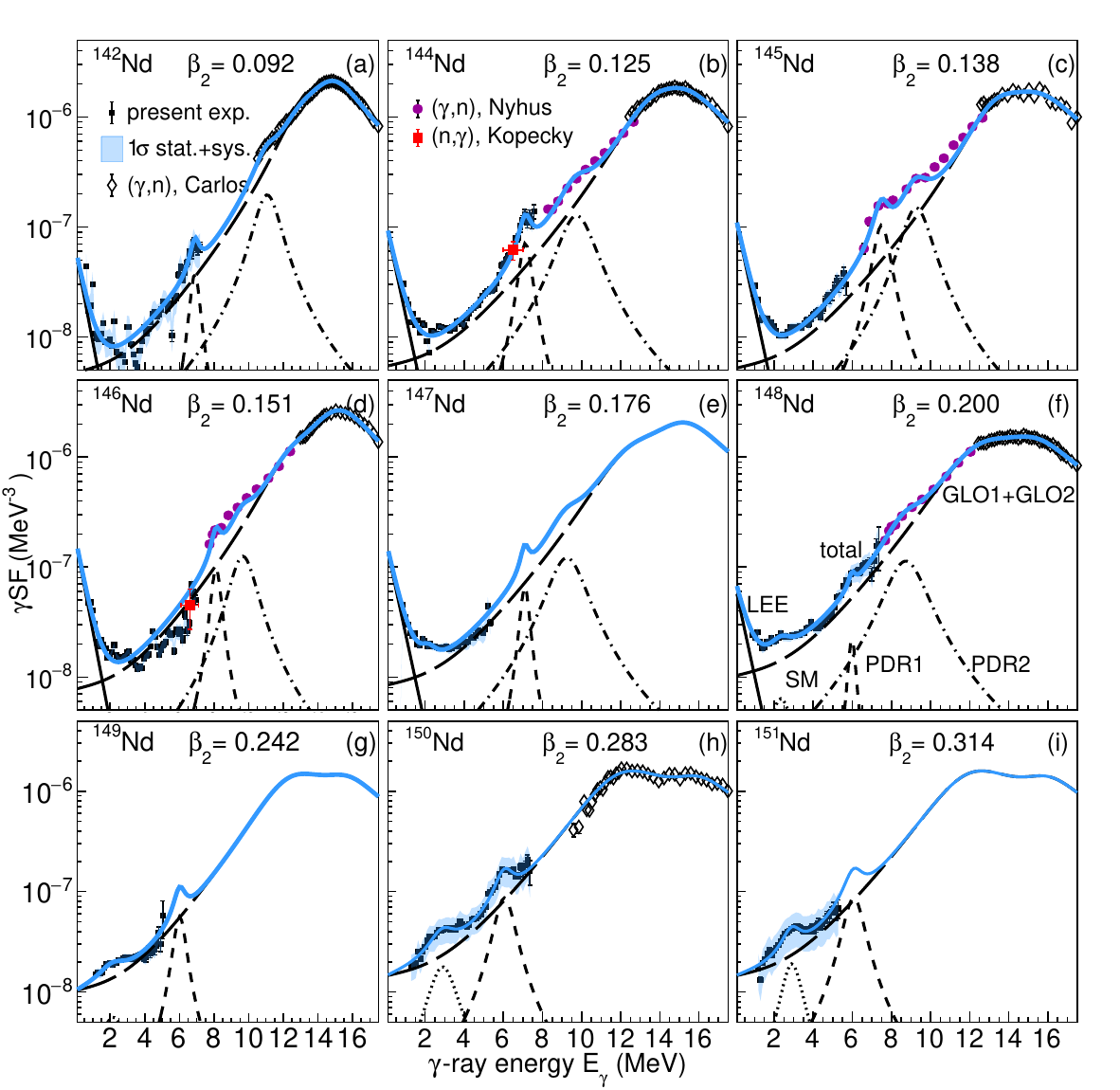}
\caption{(Color online)  Experimental $\gamma$SFs (solid squares) with its error bands (shaded blue) due to the uncertainty in $\sigma$, $D_0$, $\langle\Gamma_{\gamma 0}\rangle$ and $\eta$ parameters. The photoneutron data of Carlos~\cite{carlos1971} and Nyhus~\cite{nyhus2015} (purple circles and open diamonds) are shown for the high-energy regions. Also the resonance capture data ~\cite{IAEA2020} are shown as a filled red square data point in $^{144,146}$Nd. The various models applied are shown as curves, see assignments in panel (f). The large systematic errors for $^{150,151}$Nd are mainly due to the uncertainties in the $\left< \Gamma_\gamma\right>$ values. }
\label{fig:gsf_all}
\end{center}
\end{figure*}

We should point out that the Oslo method cannot separate the data into E1 or M1 contributions. Furthermore, the technique is restricted to an excitation energy of maximum $S_n$, which is typically around $5-6$ and $7-8$~MeV for the odd- and even-mass isotopes, respectively. Nevertheless, by also exploiting other experimental data, we will obtain a reliable description of the $\gamma$SF. Figure~\ref{fig:gsf_all} presents our data together with other external data for $^{142,144-151}$Nd. The various models are shown as curves with corresponding model parameters listed in Tables~\ref{tab:GDR} and \ref{tab:other_resonances}.

\subsection{Giant dipole resonance}

In the macroscopic picture, the GDR describes the dipole oscillations of the proton and neutron clouds against each other. The GDR is known to split into two components (GDR1 and GDR2) for nuclei with an appreciable ground-state deformation.

There exist photoneutron cross sections from the Saclay measurements on $^{142-146,148,150}$Nd by Carlos {\em et al.}~\cite{carlos1971} and from inverse Compton scattering on $^{143-146,148}$Nd by Nyhus {\em et al.}~\cite{nyhus2015}. The $\gamma$SF is calculated from the cross section by~\cite{axel1962}
\begin{equation}
f(E_{\gamma})= \frac{1}{3(\pi \hbar c)^2}\frac{\sigma_{\gamma n}(E_{\gamma})}{E_{\gamma}},
\end{equation}
where the constant reads $1/3(\pi \hbar c)^2=8.674\times 10^{-8}$mb$^{-1}$MeV$^{-2}$. The GDR data of Fig.~\ref{fig:gsf_all} show the expected spreading width~\cite{rusev2008} of $\approx 4$~MeV for the almost spherical $^{142}$Nd. For the heavier neodymiums, the GDR gets broader with deformation and finally splits into two clear GDR components for $^{150}$Nd. We also recognize for $^{144,145,148}$Nd that the high-energy Oslo data match very well with the low-energy ($\gamma,n$) data of Nyhus {\em et al.}~\cite{nyhus2015}, which give further support to the spin restrictions introduced in Sect.~III.
In the case of $^{146}$Nd, it is difficult to conclude on the degree of matching.

To describe the GDR data, we use two generalized Lorentzians (GLOs), each with the functional form of~\cite{Capote09}:
\begin{align}
&f_{\rm E1}(E_{\gamma}) = \frac{1}{3\pi^2\hbar^2c^2}\sigma_{E1}\Gamma_{E1}   \nonumber \\
                            & \times \left[ \frac{ E_{\gamma} \Gamma(E_{\gamma},T_f)}{(E_\gamma^2-E_{E1}^2)^2 + E_{\gamma}^2 \Gamma^2(E_{\gamma},T_f)}
                             + 0.7 \frac{\Gamma(E_{\gamma}=0,T_f)}{E_{E1}^3}\right]
\label{eq:GLO}
\end{align}
with
\begin{equation}
\Gamma(E_{\gamma},T_f) = \frac{\Gamma_{E1}}{E_{E1}^2} (E_{\gamma}^2 + 4\pi^2 T_f^2).
\end{equation}
The resonance parameters are the energy centroid $E_{E1}$, the strength $\sigma_{E1}$, and the width $\Gamma_{E1}$.

The $T_f$ parameter gives a non-zero tail for the lower $\gamma$ energies. The parameter is extracted from $^{142,144}$Nd that are assumed  to have no broad, smoothly-behaving features except for the LEE structure. 
We find $T_f=0.46(1)$~MeV for both isotopes, which is somewhat lower than the corresponding constant-temperature values of $T_{\rm TC}= 0.65(5)$ and $0.63(3)$~MeV. To reduce the number of free parameters in the fit, we adopt the value of $T_f=0.50(5)$~MeV for all nine isotopes.

Figure~\ref{fig:gsf_all} presents our data together with other external data for $^{142,144-151}$Nd. The E1 GLOs (long-dashed black curves) are fitted to the data of Carlos and Nyhus~\cite{carlos1971,nyhus2015}, using $T_f=0.50$~MeV. This part of the $\gamma$SFs demonstrates the interesting broadening and splitting of the GDR with deformation. For the three isotopes with no ($\gamma,n$) data, the systematics from the neighbors are adopted. In the following, we exploit the GDR data to obtain an estimate for the GDR E1 strength. The parameter sets used for modeling the GDRs are listed in Table~\ref{tab:GDR}.

\begin{table}[t]
\caption{Parameters for the GDR resonances with $T_f=0.50$~MeV. Parameters with uncertainties are from the fit.}
\begin{tabular}{lcccccc}
\hline
\hline
Nucleus&\multicolumn{3}{c}{GDR1}&\multicolumn{3}{c}{GDR2}\\
\cline{2-7}
               &$E_{\rm GDR1}$ &$\sigma_{\rm GDR1}$  &$\Gamma_{\rm GDR1}$  &$E_{\rm GDR2}$  &$\sigma_{\rm GDR2}$  &$\Gamma_{\rm GDR2}$ \\
               &(MeV)         &      (mb)     &      (MeV)    &(MeV)          &       (mb)    &     (MeV)    \\
\hline

$^{142}$Nd     &  13.5(3)   &    89(18)  &      3.3(8) &     15.3(4) &    325(22)    &    3.8(1)  \\
$^{144}$Nd     &  14.4(7)   &   179(93)  &      4.4(11)&     15.9(2) &    184(103)   &    4.3(7)  \\
$^{145}$Nd     &  14.0(4)   &   166(48)  &      3.7(9) &     16.1(2) &    209(48)    &    4.8(9)  \\
$^{146}$Nd     &  13.1(7)   &   89(19)   &      3.9(19)&     15.7(5) &    435(30)    &    4.6(2)  \\
$^{147}$Nd     &  13.5      &   130      &        5.2  &     15.8    &    292        &    4.6  \\
$^{148}$Nd     &  13.8(9)   &  172(32)   &     6.4(28) &     15.9(8) &   147(58)     &    4.6(11) \\
$^{149}$Nd     &  13.6      &   193      &         6.0 &     16.2    &    146        &    4.3  \\
$^{150}$Nd     &  13.4(2)   &   213(5)   &      6.7(5) &     16.5(8) &     145(9)    &    4.0(4)  \\
$^{151}$Nd     &  13.4      &   213      &         6.7 &      16.5   &     145       &    4.0  \\
\hline
\hline
\end{tabular}
\\
\label{tab:GDR}
\end{table}


\begin{table*}[t]
\caption{Parameters for the PDR, SM and LEE structures and the integrated $B(M1)$ for LEE and SM. A common slope parameter of $\kappa_{\rm LEE}=1.9$~MeV$^{-1}$ is set for all isotopes. Parameters with statistical and systematical uncertainties are from the fit.}
\begin{tabular}{lcccccccccccc}
\hline
\hline
Nucleus&\multicolumn{3}{c}{PDR1}&\multicolumn{3}{c}{PDR2}&\multicolumn{4}{c}{SM}&\multicolumn{2}{c}{LEE}\\
\cline{2-13}
               &$E_{\rm PDR1}$ &$\sigma_{\rm PDR1}$  &$\Gamma_{\rm PDR1}$  &$E_{\rm PDR2}$  &$\sigma_{\rm PDR2}$  &$\Gamma_{\rm PDR2}$ &$E_{\rm SM}$  &$\sigma_{\rm SM}$  &$\Gamma_{\rm SM}$&$\sum B_{\rm SM}$&$C_{\rm LEE}$  &$\sum B_{\rm LEE}$\\
               &(MeV)         &      (mb)     &      (MeV)    &(MeV)          &       (mb)    &     (MeV)    &(MeV)          &       (mb)    &     (MeV) &      ($\mu^2_N$)  &    ($10^{-8}$MeV$^{-1}$)    &     ($\mu^2_N$)\\
\hline

$^{142}$Nd  &  6.90     &   3.0      &    0.6     &  11.1     &      25.0 &   1.6     &           &          &         &       & 6.0(16) &   7.3(19)\\
$^{144}$Nd  &  7.13(9)  &   5.8(9)   &  0.71(14)  &  9.80(10) &  14.3(12) & 1.96(12)  &           &          &         &       &10.5(11) &  14.1(15)\\
$^{145}$Nd  &  7.5      &    9.0     &   0.9      &  9.3      &     16    &  1.6      &           &          &         &       &12.6(21) &  16.4(27)\\
$^{146}$Nd  &  8.12(2)  &   9.3(4)   &   0.6      &  9.7      &      14   &  1.6      &           &          &         &       &17.5(27) &  21.4(33)\\
$^{147}$Nd  &  7.1      &    5.5     &   0.5      &  9.3      &       13  &  2.0      & 2.22(16)  & 0.12(4)  &  1.0(5) &1.6(10)&16.5(11) &  20.1(13)\\
$^{148}$Nd  &  5.98(13) &   1.6(9)   & 0.39(26)   &  8.80(9)  &   11.5(8) &2.45(13)   & 2.45(14)  & 0.18(4)  &  1.4(5) &2.8(12)&6.9(12)  &   9.0(16)\\
$^{149}$Nd  &  6.03     &    3.9     &  0.75      &           &           &           &  2.37(16) & 0.15(4)  &  1.9(5) &3.1(12)&         &          \\
$^{150}$Nd  &  6.08(17) &    6.1(18) & 1.1(4)     &           &           &           &3.00(30)   & 0.61(18) &  1.5(7) &8.5(46)&         &          \\
$^{151}$Nd  &   6.1     &    6.1     & 1.1        &           &           &           &2.95(25)   & 0.64(27) &  1.1(6) &7.0(47)&         &          \\

\hline
\hline
\end{tabular}
\\
\label{tab:other_resonances}
\end{table*}

\subsection{Pygmy dipole resonance}

The pygmy dipole resonance (PDR) is believed to originate from oscillations of the neutron skin against the $N=Z=60$ core. The number of excess neutrons $\Delta N = A-2Z$  changes from $\Delta N = 22$ for $^{142}$Nd to $\Delta N = 31$ for $^{151}$Nd, and the strength of the PDR resonance is expected to increase with this neutron excess. An overview of experimental approaches to study the low-lying electric dipole strength and experimental results are given by Savran {\em et al.}~\cite{savran2013}.

The resonance is described by the standard Lorentzian (SLO) model~\cite{Capote09} given by
\begin{equation}
f(E_{\gamma}) = \frac{1}{3\pi^2\hbar^2c^2}\frac{\sigma E_{\gamma} \Gamma^2}{(E_\gamma^2-E^2)^2 + E_{\gamma}^2 \Gamma^2},
 \label{eq:SLO}
\end{equation}
with parameters $E$, $\sigma$, and $\Gamma$ appropriate for the specific resonance.

By inspecting Fig.~\ref{fig:gsf_all}, we find resonance structures in the energy regions of $6 - 8$~MeV and $9 - 11$~MeV that we for simplicity denote PDR1 and PDR2, respectively. The lower resonance PDR1 is probably the one traditionally called the pygmy dipole resonance. The high-lying resonance PDR2, which accounts for the strength in the $\gamma$-energy region starting around 11.3 MeV for $^{142}$Nd and decreasing with mass number to 8.7 MeV for $^{148}$Nd, may as well be due to a fragmentation of the GDR. Reinhard and Nazarewicz~\cite{reinhard2013} have used nuclear density functional theory to describe a weak collective E1 structure in the region below the GDR. Thus, the proper interpretation of the observed PDR2 is uncertain.

The most clear fingerprints for the PDR1 are found in the $^{145}$Nd data of Nyhus {\em et al.}~\cite{nyhus2015} and in the present $^{148,150}$Nd Oslo data. For $^{145}$Nd, the PDR1 is located at $E_{\gamma}\approx7.5$~MeV and drops to $\approx 6.0$~MeV for $^{148,150}$Nd. However, there is no clear evidence in the compiled data sets that the energy centroids decrease in a smooth way with increasing mass number. Also, the evolution of the PDR1 strength seems complicated as it has a minimum for the transitional $^{148}$Nd isotope.

For $^{147,149,151}$Nd there are data of neither the PDRs nor the GDRs, but we include these structures in the over-all fit in order to get a consistent description of the underlying strength of the low-energy LEE and SM structures. For these isotopes, the resonance parameters are estimated from the neighboring isotopes with values listed in Tables~\ref{tab:GDR} and \ref{tab:other_resonances}.

\subsection{Low-energy enhancement and scissors mode}
The {\em low-energy enhancement} (LEE), often called upbend, was first observed in $^{56}$Fe~\cite{voinov2004} and then in $^{93-98}$Mo~\cite{guttormsen2005}. The phenomenon manifests itself as an increase in $\gamma$ strength with decreasing energy below $E_{\gamma}\approx3$~MeV. The LEE, which is embedded in the nuclear quasicontinuum, was completely unexpected at that time. Eight years later, Wiedeking~{\em et al.}~\cite{wiedeking2012} confirmed the same structure in $^{95}$Mo using other detectors and techniques. The LEE has been found for many lighter nuclei, but has recently also been observed in heavier isotopes like $^{138,139}$La~\cite{kheswa2015}, $^{147,149}$Sm~\cite{naqvi2019} and $^{151,153}$Sm~\cite{simon2016}. 
However, more recent experiments on $^{153,155}$Sm~\cite{malatji2021} did not reveal any LEE. This makes it questionable if LEE exists also for heavy deformed systems. 

The first modeling of the LEE structure was performed by Litvinova and Belov in 2013~\cite{litvinova2013} within finite-temperature mean-field theory, adding a temperature parameter to explain the LEE in terms of an increased $E1$ component.
The same year, Schwengner {\em et al.}~\cite{schwengner2013} published a comparison with shell-model calculations of the $M1$ strength function and experimental data for $^{90}$Zr and $^{94-96}$Mo. 
They suggested an exponential form of the LEE given by
\begin{equation}
f(E_{\gamma}) = C \exp (-\kappa E_{\gamma}),
 \label{eq:LEE}
\end{equation}
where $C$ and $\kappa$ are parameters.
Recently, Midtb{\o} {\em et al.}~\cite{midtbo2018} reviewed the present status of the LEE. They reported on a systematic large-scale shell-model study of 283 nuclei, which reveals in general a more steep and pronounced LEE as the mass number increases.

In this work, we follow the suggested description of Ref.~\cite{schwengner2013} to model the LEE.
The slope parameter $\kappa$ of Eq.~(\ref{eq:LEE}) is determined by fits to the pronounced LEE structures of $^{142,144-147}$Nd shown in Fig.~\ref{fig:gsf_all}, which give an average value of $\kappa = 1.9(2)$~MeV$^{-1}$. The $\kappa$ parameter may be interpreted as the inverse of temperature~\cite{schwengner2013}. However,  $1/1.9\approx 0.53$~MeV do not coincide with the average value of $T_{\rm CT}\approx 0.61$~MeV for $^{142,144-147}$Nd, see Table~\ref{tab:gsf_parameters}. Therefore, we use a fixed value of $\kappa = 1.9(2)$~MeV$^{-1}$ in the fits for the $^{142,144-148}$Nd isotopes. The fitted values of the remaining $C$ parameter are listed in Table~\ref{tab:other_resonances}. For isotopes heavier than $^{148}$Nd, we see no clear LEE structure with the present experimental lower $\gamma$-energy threshold. Variations in the $T_f$ value (from 0.45 to 0.55 MeV) are found to give only minor changes in the extracted LEE strength.

The {\em scissors mode} (SM), which appears in deformed nuclei, was already described in 1978 by Lo Iudice and Palumbo~\cite{iudice1984}. 
They proposed a geometrical picture where the deformed proton and neutron clouds oscillate against each other like scissors blades.
Inspired by these predictions, the ($e,e'$) reaction, which probes transition strengths from the ground-state to excited levels in the low excitation-energy region, was used to reveal M1 type resonant levels in $^{156,158}$Gd~\cite{bohle1984}. 
It is interesting that these authors also report on the absence of such levels in $^{146}$Nd, which is highly relevant for the present work. 
Furthermore, nuclear resonance fluorescence (NFR) experiments~\cite{peter1995} reveal SM $\gamma$-transition strengths not only to the ground-state, but also eventually to the very lowest excited levels.
An alternative prediction of strong low-energy transitions was given in 1982 by Chen and Leander\cite{chen1982}, which involves $\gamma$ decay in the nuclear quasicontinuum. Their calculations show that strong $M1$ transitions between $\Delta \Omega = 1$ Nilsson orbitals\footnote{The Nilsson orbitals are assigned a quantum number $\Omega$, which represents the projection of {\bf $j$} on the nuclear symmetry axis.} could compete with the statistical decay in the quasicontinuum. These predictions were confirmed in 1984 by the observation of a bump at $E_{\rm SM}\approx 2-3$~MeV in the $\gamma$-ray spectra from various excitation energies in the quasicontinuum of $^{161}$Dy~\cite{guttormsen1984}. 
It has been shown that the SM in the low excitation-energy region increases in strength with deformation~\cite{peter1995}. 
A comprehensive review of magnetic dipole excitations is given by Heyde {\em et al.}~\cite{heyde2010}.
The main techniques used to study the SM on excited states in the quasicontinuum are the Oslo method applied in the present work, and the multistep cascade (MSC) method, see e.g. Ref.~\cite{valenta2017}.

In the present work we study the SM embedded in the nuclear quasicontinuum applying the Oslo method, see e.g. the study of $^{160-164}$Dy by Renstr{\o}m {\em et al.}~\cite{renstrom2018} and references therein. In deformed rare earth nuclei, the SM follows closely a standard Lorentzian at $E_{\rm SM}\approx 2-3$~MeV. In the actinide region, the SM strength increases and is best modeled by two close-lying Lorentzians~\cite{tornyi2014,guttormsen2014}.

Figure~\ref{fig:gsf_all}(e) indicates a weak SM at $E_{\rm SM}\approx2.2$~MeV in $^{147}$Nd, which results in a rather flat $\gamma$SF located in-between the LEE structure and the tail of the GDR. It is interesting to observe how the $\gamma$SF at these energies grows in strength with deformation and finally reveals a clear and strong SM structure in $^{150,151}$Nd at $E_{\rm SM}\approx3.0$~MeV.

Table~\ref{tab:other_resonances} shows that the SM energy centroid and strength resonance parameters vary in a possibly systematic way. The clear exception is the width of $^{149}$Nd. There is a possibility that there is a significant LEE strength in $^{149}$Nd, comparable to $^{148}$Nd. If so, this would reduce the width of the SM. As we have not included an LEE in the fit, as our data are insufficient to conclude on its existence in this case, we cannot say whether this is the main reason or not. The $^{149}$Nd nucleus is also dependent on interpolations from the neighboring nuclei. This may cause the large resonance width of 1.9~MeV from the fit, or that this nucleus for some structural reason really exhibits a large width.

\begin{figure}[ht]
\begin{center}
\includegraphics[clip,width=1.\columnwidth]{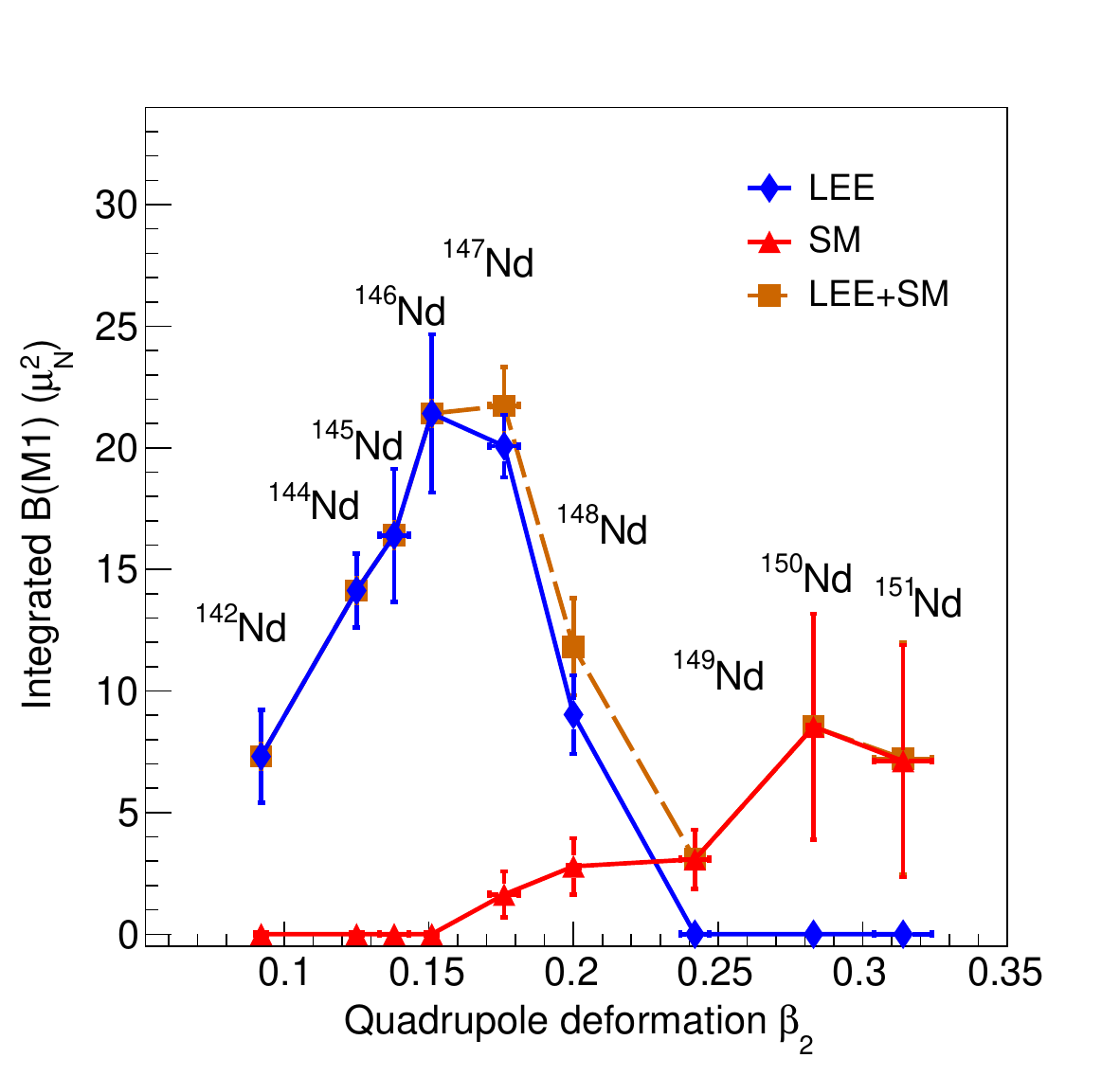}
\caption{(Color online) The integrated LEE (blue diamonds) and SM (red triangles) strengths $B(M1)$ together with the summed strength of the two structures (orange squares) as function of quadrupole deformation $\beta_2$. The strengths are integrated between $E_{\gamma}=0$ and $5$~MeV.}
    \label{fig:LEESR}
\end{center}
\end{figure}

An intriguing idea was proposed by Schwengner {\em et al.}~\cite{schwengner2017} based on large-scale shell-model calculations for $^{60,64,68}$Fe. They find that the strength of the SM increases by a factor of two when going from $^{60}$Fe to $^{68}$Fe. At the same time, the LEE strength decreases correspondingly and thereby conserves the total strength of $B(M1)\approx 9.8 \mu_N^2$. The conservation of the summed SM and LEE strengths has been experimentally tested for $^{147,149,151,153}$Sm~\cite{naqvi2019}, but large uncertainties prohibit a firm conclusion.

The evolution of the LEE and SM is shown in Fig.~\ref{fig:gsf_all}. It is clear that the LEE is present in $^{142,144-148}$Nd and the SM is present in $^{147-151}$Nd. Here, $^{148}$Nd is a key nucleus since both the LEE and SM are significantly present. 
For $^{145,146}$Nd, we do not have sufficient experimental evidence from our data to claim the presence of the scissors mode. It is interesting to note that no SM strength could be seen for $^{146}$Nd in the ($e,e'$) experiments~\cite{bohle1984}.

In order to quantify the strength of the LEE and SM resonances as function of deformation (or mass number), we integrate the corresponding $\gamma$SFs for the two structures. 
The $\gamma$-decay strength from an initial level with spin $J_i$ is proportional with the number of available final spins $J_f$ that can be reached with a transition of electromagnetic character $XL$. If we further assume an initial spin with $J_i\geq L$, we obtain~\cite{Bartholomew1973,Lone1986}
\begin{align}
    \frac{dB(XL)}{dE_{\gamma}} &= \frac{L \left[ (2L+1)!!\right]^2{\left({\hbar c}\right)^{2L+1}}}{8\pi(L+1)} f_{XL}(E_\gamma)\sum_{J_f}1 \nonumber\\
    &= \frac{L\left[ (2L+1)!!\right]^2{\left({\hbar c}\right)^{2L+1}}}{8\pi(L+1)} f_{XL}(E_\gamma)(2L+1).
\end{align}
This expression is now used to evaluate the total strength in the $\gamma$-energy region $0-5$~MeV. Assuming M1 electromagnetic character, the upward integrated strength with $L=1$ is given by
\begin{equation}
B(M1)=\frac{27(\hbar c)^3}{16\pi}\int_{0}^{\rm 5 MeV} f(E_{\gamma}) dE_{\gamma},
 \label{eq:bm1}
\end{equation}
where $f(E_{\gamma})$ is modeled by Eqs.~(\ref{eq:SLO}) and (\ref{eq:LEE}) for SM and LEE, respectively. The factor in front of the integral has the value $27(\hbar c)^3/16\pi= 2.598\times10^8 \mu^2_N$MeV$^2$.

The LEE and SM strengths can now be evaluated from their respective fit functions shown in Fig.~\ref{fig:gsf_all} using Eq.~(\ref{eq:bm1}).
Figure~\ref{fig:LEESR} summarizes the total M1 strengths for the nine isotopes of the present experiment, which are also included in Table~\ref{tab:other_resonances}. The error bars are mainly due to uncertainties in the $T_f=0.50(5)$ parameter and the experimental $\langle \Gamma_{\gamma} \rangle$ value. 

The blue curve of Fig.~\ref{fig:LEESR} shows the rise and fall of the LEE strength with a maximum of $\approx 21 \mu^2_N$ for $^{146,147}$Nd with $\beta_2 \approx 0.16$. For $^{148}$Nd the strength drops to $\approx 9 \mu^2_N$ and then vanishes for the heavier isotopes. The strength of the SM (red curve) starts increasing at $^{147}$Nd and reaches a plateau of $\approx 8 \mu^2_N$ for the well-deformed $^{150,151}$Nd. 

The summed strength of the LEE and SM structures is shown as an orange curve in Fig.~\ref{fig:LEESR}.
It is clear that this curve is far from constant and contradicts the picture of Schwengner {\em et al.}~\cite{schwengner2017}. Firstly, we find that the SM structure does not account for the missing strength of the LEE at $^{149-151}$Nd; it reaches only one third of the maximum LEE strength. Secondly, there is a strong increase of LEE strength from $^{142}$Nd to $^{146,147}$Nd where no SM strength is present that could eventually account for this behavior. 

The $^{148}$Nd seems to be a key nucleus where both the LEE and SM are coexisting. It is noteworthy that the clear onset of the SM structure for this transitional nucleus coincides with the onset of collectivity in the mean-field solution for the shell-model interaction~\cite{Gut2021}.

\section{Conclusions and outlook}

The present study has shown that the limitation of transferred spin in the present ($p,p'$) and the ($d,p$) reactions has to be taken into account when using the Oslo method. 
By including the spin reduction, the slope of the $\gamma$SF is reduced.
The applied spin corrections are supported by the shape method and $\gamma$-ray side feeding into the rotational ground-state band. The corrected $\gamma$SF also matches ($n, \gamma$) and ($\gamma,n$) data available from literature.

The ($\gamma,n$) data on the giant dipole resonances are modeled using the GLO model with a fixed $T_f = 0.50(5)$~MeV. The fitted resonance parameters are exploited to obtain the E1 GDR tail, which is underlying the $\gamma$SF structures located at lower $\gamma$ energies. 

The pygmy dipole resonance is partly obtained from the present Oslo data and the ($\gamma,n$) data from literature. The data of $^{142,144-146,148}$Nd indicate the presence of two resonances. However, we suggest that the one lower in energy is related to the pygmy dipole resonance built on neutron skin oscillations.

The present Oslo data bring new insight into the evolution of the low-energy enhancement (LEE) and the scissors mode (SM). By increasing deformation, we observe around $^{148}$Nd with $\beta_2 \approx 0.2$ that the LEE strength decreases and the SM strength increases. Apart from that, there seems not to be any connection between the amount of $B(M1)$ strengths carried by the two structures as function of deformation. This contradicts the idea that the LEE and SM structures~\cite{schwengner2017,naqvi2019,frauendorf2022} are connected by exchanging strength with each other. Thus, the evaluated $B(M1)$ strengths indicate that the two structures are due to different mechanisms.

In conclusion, the present work reveals the fascinating evolution of various $\gamma$-ray structures in the cross over from spherical to deformed neodymium isotopes. There is a clear need for more data and in particular, more robust theoretical descriptions to understand the interplay between $\gamma$-decay mechanisms in the nuclear quasicontinuum.

\begin{acknowledgments}
We thank J.~C.~M\"uller, P.~Sobas and J.~Wikne for providing excellent experimental conditions. We gratefully acknowledge valuable discussions with  Sean Liddick, Dennis M{\"u}cher, Artemis Spyrou and Fabio Zeiser. This work was supported by The Scientific and Technological Research Council of Turkey (TUBITAK) with Project No.~115F196. A.~C.~L. acknowledges funding of this research by the European Research Council through ERC-STG-2014, Grant Agreement No.~637686.
This work was partially supported by Projects No.~263030 and No.~262952 of the Norwegian Research Council. The OSCAR detector was funded by the Norwegian Research Council Project No.~245882. This work is based on the research supported in part by the National Research Foundation of South Africa (Grant No.~118846).
\end{acknowledgments}


\begin{thebibliography}{99}

\bibitem{Bartholomew1973} G.~A.~Bartholomew, E.~D.~Earle, A.~J.~Fergusson, J.~W.~Knowles and M.~A.~Lone, Adv. Nucl. Phys. {\bf 7}, 229 (1973).
\bibitem{Goriely2019} S.~Goriely {\em et al.}, Eur. Phys. J. A {\bf 55}, 172 (2019).
\bibitem{IAEA2020}  IAEA Nuclear Data Services, \texttt{htpps://www-nds.iaea.org/PSFdatabase}.
\bibitem{schwengner2017} R.~Schwengner, S.~Frauendorf, and B.A. Brown, Phys. Rev. Lett. {\bf 118}, 092502 (2017).
\bibitem{naqvi2019} F. Naqvi {\em et al.}, Phys.\ Rev.\ C\bf 99\rm, 054331 (2019).
\bibitem{frauendorf2022} S. Frauendorf and R. Schwengner, Phys.\ Rev.\ C\bf 105\rm, 034335 (2022).
\bibitem{savran2013} D. Savran, T. Aumann, and A. Zilges, Prog. Part. Nucl. Phys. {\bf 70}, 210 (2013).
\bibitem{Gut87} M.~Guttormsen, T.~Rams{\o}y, and J.~Rekstad, Nucl.\ Instrum.\ Methods Phys.\ Res.\ A \bf 255\rm, 518 (1987).
\bibitem{Gut96} M.~Guttormsen, T.~S.~Tveter, L.~Bergholt, F.~Ingebretsen, and J.~Rekstad, Nucl.\ Instrum.\ Methods Phys.\ Res.\ A \bf 374\rm, 371 (1996).
\bibitem{Schiller00} A.~Schiller, L.~Bergholt, M.~Guttormsen, E.~Melby, J.~Rekstad, and S.~Siem, Nucl. Instrum. Methods Phys. Res. A {\bf 447} 494 (2000).
\bibitem{Wiedeking2021} M.~Wiedeking, M.~Guttormsen, A.~C.~Larsen, F.~Zeiser, A.~G\"{o}rgen, S.~N.~Liddick, D.~M\"{u}cher, S.~Siem, and A.~Spyrou. Phys.\ Rev.\ C \bf 104\rm, 014311 (2021).
\bibitem{Gut2021} M.~Guttormsen {\em et al.}, Phys.\ Lett.\ B \bf 816\rm, 136206 (2021) and supplemantary material.
\bibitem{siri} M.~Guttormsen, A.~B\"urger, T.~E.~Hansen, and N.~Lietaer, Nucl.~Instrum.~Methods Phys.~Res.~A {\bf 648}, 168 (2011).
\bibitem{CACTUS} M.~Guttormsen, A.~Atac, G.~L{\o}vh{\o}iden, S.~Messelt, T.~Rams{\o}y, J.~Rekstad, T.~F.~Thorsteinsen, T.S.~Tveter, and Z.~Zelazny, Phys.\ Scr. \bf T 32\rm, 54 (1990).
\bibitem{Zeiser2021} F.~Zeiser {\em et al.}, Nucl. Instr. Methods A {\bf985}, 164678 (2021).
\bibitem{Goergen2021} A.~G\"{o}rgen, M.~Guttormsen, A.~C.~Larsen, S.~Siem, E.~Adli, N.~F.~J.~Edin, H.~Gjerstad, G.~Henriksen, E.~Malinen, V.~Modamio, B.~Schoultz, P.~A.~Sobas, T.~A.~Theodossiou and J.~C.~Wikne, Eur. Phys. J. Plus {\bf 136}, 181 (2021).
\bibitem{om2022} M.~Guttormsen, A.~C.~Larsen, F.~Zeiser, J.~E.~Midtb\o{} and V.~W.~ Ingeberg, Oslo Method Software v1.1.6 (2022). Available on-line at \texttt{https://github.com/oslocyclotronlab}.
\bibitem{dirac} P.~A.~M.~Dirac, Proc. R. Soc. Lond. {\bf A 114}, 243 (1927).
\bibitem{fermi} E.~Fermi, Nuclear Physics (University of Chicago Press, Chicago, 1950).
\bibitem{brink} D.~M.~Brink, Doctorial thesis, Oxford University, 1955 (unpublished).
\bibitem{guttormsen2016} M.~Guttormsen, A.~C.~Larsen, A.~G{\"o}rgen, T.~Renstr{\o}m, S.~Siem, T.~G.~Tornyi and G.~M.~Tveten, Phys. Rev. Lett. {\bf 116}, 012502 (1990).
\bibitem{kopecky1990} J. Kopecky and M. Uhl, Phys.\ Rev.\ C \bf 41\rm, 1941 (1990).
\bibitem{Capote09} R.~Capote, M.~Herman, P.~Oblozinsky, {\em et al.}, Nuclear Data Sheets {\bf 110}, 3107 (2009). Reference Input Library RIPL-3  available online at {\it http://www-nds.iaea.org/RIPL-3/}.
\bibitem{NNDC} Data from the NNDC On-Line Data Service database; available at \texttt{http://www.nndc.bnl.gov/nudat2/}.
\bibitem{MugAtlas} S.~F.~Mughabghab, {\em Atlas of Neutron Resonances}. (Elsevier Science, Amsterdam, 2018). 6th ed.
\bibitem{Ericson59}  T.~Ericson, Nucl. Phys.  {\bf 11}, 481(1959).
\bibitem{pritychenko2016} B.~Pritychenko {\em et al.} Atomic Data and Nuclear Data Tables {\bf 107}, 1–139 (2016).
\bibitem{voin1} A.~Voinov, M.~Guttormsen, E.~Melby, J.~Rekstad, A.~Schiller, and S.~Siem, Phys.\ Rev.\ C \bf 63\rm, 044313 (2001).
\bibitem{Lars11} A.~C.~Larsen {\em et al.}, Phys.\ Rev.\ C \bf 83\rm, 034315 (2011).
\bibitem{james1981} F.James. Determining the statistical significance of experimental results, CERN Technical Reports No. DD/81/02 and No. 81-03, 1981 (unpublished).
\bibitem{carlos1971} P.~Carlos, H.~Beil, R.~Bergere, A.~Lepretre and A.~Veyssiere, Nucl.\ Phys.\ A \bf 172\rm, 437 (1971).
\bibitem{nyhus2015} H.-T.~Nyhus {\em et al.}, Phys.\ Rev.\ C \bf 91\rm, 015808 (2015).
\bibitem{axel1962} P. Axel, Phys. Rev. {\bf 126}, 671 (1962).
\bibitem{rusev2008} G. Rusev {\em et al.}, Phys.\ Rev.\ C \bf 77\rm, 064321 (2008).
\bibitem{reinhard2013} P.-G. Reinhard and W. Nazarewicz,  Phys.\ Rev.\ C \bf 87\rm, 014324 (2013).
\bibitem{voinov2004} A. Voinov, E. Algin, U. Agvaanluvsan, T. Belgya, R. Chankova, M. Guttormsen, G.E. Mitchell, J. Rekstad, A. Schiller, and S. Siem, Phys.\ Rev.\ Lett. \bf 93\rm, 142504 (2004).
\bibitem{guttormsen2005} M. Guttormsen {\em et al.}, Phys.\ Rev.\ C \bf 71\rm, 044307 (2005).
\bibitem{wiedeking2012} M. Wiedeking {\em et al.}, Phys.\ Rev.\ Lett.\bf 108\rm, 162503 (2012).
\bibitem{kheswa2015} B. V. Kheswa {\em et al.}, Phys.\ Lett. B\bf 744\rm, 268 (2015).
\bibitem{simon2016} A. Simon {\em et al.}, Phys.\ Rev.\ C\bf 93\rm, 034303 (2016).
\bibitem{malatji2021} K. L. Malatji {\em et al.}, Phys.\ Rev.\ C\bf 103\rm, 014309 (2021).
\bibitem{litvinova2013} Elena Litvinova and Nikolay Belov, Phys. Rev. C \textbf{88}, 031302 (R) (2013).
\bibitem{schwengner2013} R.~Schwengner, S.~Frauendorf, and A.~C.~Larsen, Phys. Rev. Lett. {\bf 111}, 232504 (2013).
\bibitem{midtbo2018} J.E. Midtb{\o}, A.C. Larsen, T. Renstr{\o}m, F.L. Bello Garrote, and E. Lima, Phys.\ Rev.\ C\bf 98\rm, 064321 (2018).
\bibitem{iudice1984} N. Lo Iudice and F. Palumbo, Phys. Rev. Lett. {\bf 41}, 1532 (1978).
\bibitem{bohle1984} D. Bohle, A. Richter, W. Steffen A.E.L.Dieperink, N. Lo Iudice, F. Palumbo, and O. Scholte, Phys. Lett. B {\bf 137} 27 (1984).
\bibitem{peter1995} P. von Neumann-Cosel, J.N. Ginocchio, H. Bauer,  and A. Richter, Phys. Rev. Lett. {\bf 75}, 23 (1995).
\bibitem{chen1982} Y. S. Chen and G. A. Leander, Phys. Rev. C {\bf 26}, 2607 (1982).
\bibitem{guttormsen1984} M. Guttormsen, J. Rekstad, A. Henriquez, F. Ingebretsen, and T.F. Thorsteinsen, Phys. Rev. Lett. {\bf 52}, 102 (1984).
\bibitem{heyde2010} K. Heyde, P. von Neumann-Cosel, and A. Richter, Rev. Mod. Phys. {\bf 82}, 2365 (2010).
\bibitem{valenta2017} S.~Valenta {\em et al.}, Phys.\ Rev.\ C\bf 96\rm, 054315 (2017).
\bibitem{renstrom2018} T. Renstr{\o}m {\em et al.}, Phys.\ Rev.\ C\bf 98\rm, 054310 (2018).
\bibitem{tornyi2014} T. G. Tornyi {\em et al.}, Phys.\ Rev.\ C\bf 89\rm, 044323 (2014).
\bibitem{guttormsen2014} M. Guttormsen {\em et al.}, Phys.\ Rev.\ C\bf 89\rm, 014302 (2014).
\bibitem{Lone1986} M.~A. Lone. Neutron Induced Reactions, eds.~ J. Kri\u{s}tiak and E.~B\u{e}t\'{a}k, Springer, Dordrecht (1986) p.238-252.

\end{thebibliography}
\end{document}